\documentclass[epj]{svjour}
\pdfoutput=1
\RequirePackage{snapshot} % required by bundledoc
\usepackage[utf8]{inputenc}
\usepackage[merge,numbers,compress]{natbib}
\usepackage{bm,amsmath}
\usepackage{amssymb}
\usepackage{graphicx}
\usepackage{subfig}
\usepackage[draft]{hyperref}
\usepackage{xspace}
\usepackage{wasysym} %For permil sign
\usepackage{epstopdf}
\usepackage{booktabs}
\usepackage{cuted}
\usepackage{microtype}
\usepackage{color} % for comments
\usepackage{amssymb}
\usepackage{amsmath}

\usepackage{verbatim}

\AtBeginDocument{
\heavyrulewidth=.08em
\lightrulewidth=.05em
\cmidrulewidth=.03em
\belowrulesep=.65ex
\belowbottomsep=0pt
\aboverulesep=.4ex
\abovetopsep=0pt
\cmidrulesep=\doublerulesep
\cmidrulekern=.5em
\defaultaddspace=.5em
}

\newcommand{\as}{\alpha_s}

\newcommand{\MeV}{\;\mathrm{MeV}}
\newcommand{\GeV}{\;\mathrm{GeV}}
\newcommand{\TeV}{\;\mathrm{TeV}}
\newcommand{\muR}{\mu_{\mathrm{R}}}
\newcommand{\muF}{\mu_{\mathrm{F}}}

\newcommand{\beq}{\begin{equation}}
\newcommand{\eeq}{\end{equation}}

\newcommand{\vbone}{ \bm{q}_1 }
\newcommand{\vbtwo}{ \bm{q}_2 }

\newcommand{\ed}{\end{document}}

\begin{document}\sloppy
\title{On the impact of non-factorisable corrections in VBF single and double Higgs production}

\newcommand{\OXaff}{Rudolf Peierls Centre for Theoretical Physics,
  University of Oxford,
  Clarendon Laboratory, Parks Road, Oxford OX1 3PU}

\author{Fr\'ed\'eric A. Dreyer \and Alexander Karlberg \and Lorenzo Tancredi}
\institute{\OXaff}

\abstract{%
  We study the non-factorisable QCD corrections, computed in the
  eikonal approximation, to Vector-Boson Fusion single and double
  Higgs production and show the combined factorisable and
  non-factorisable corrections for both processes at
  $\mathcal{O}(\as^2)$.
  We investigate the validity of the eikonal approximation with and
  without selection cuts, and carry out an in-depth study of the
  relative size of the non-factorisable next-to-next-to-leading
    order corrections compared to the factorisable ones.
  After selection cuts are
  applied, the non-factorisable corrections are found to be mostly
  contained within the factorisable scale uncertainty bands.  When no
  cuts are applied, instead, the non-factorisable corrections are
  slightly outside the scale uncertainty band.
  All contributions studied here have been implemented in {\tt proVBFH}
  v1.2.0 and {\tt proVBFHH} v1.2.0.
  }

\date{\today}
\journalname{OUTP-20-02P}
\maketitle

%----------------------------------------------------------------------
\section{Introduction}
\label{sec:intro}

\begin{figure*}[h!]
  \centering
  \subfloat[Factorisable corrections]{\includegraphics[width=0.5\textwidth]{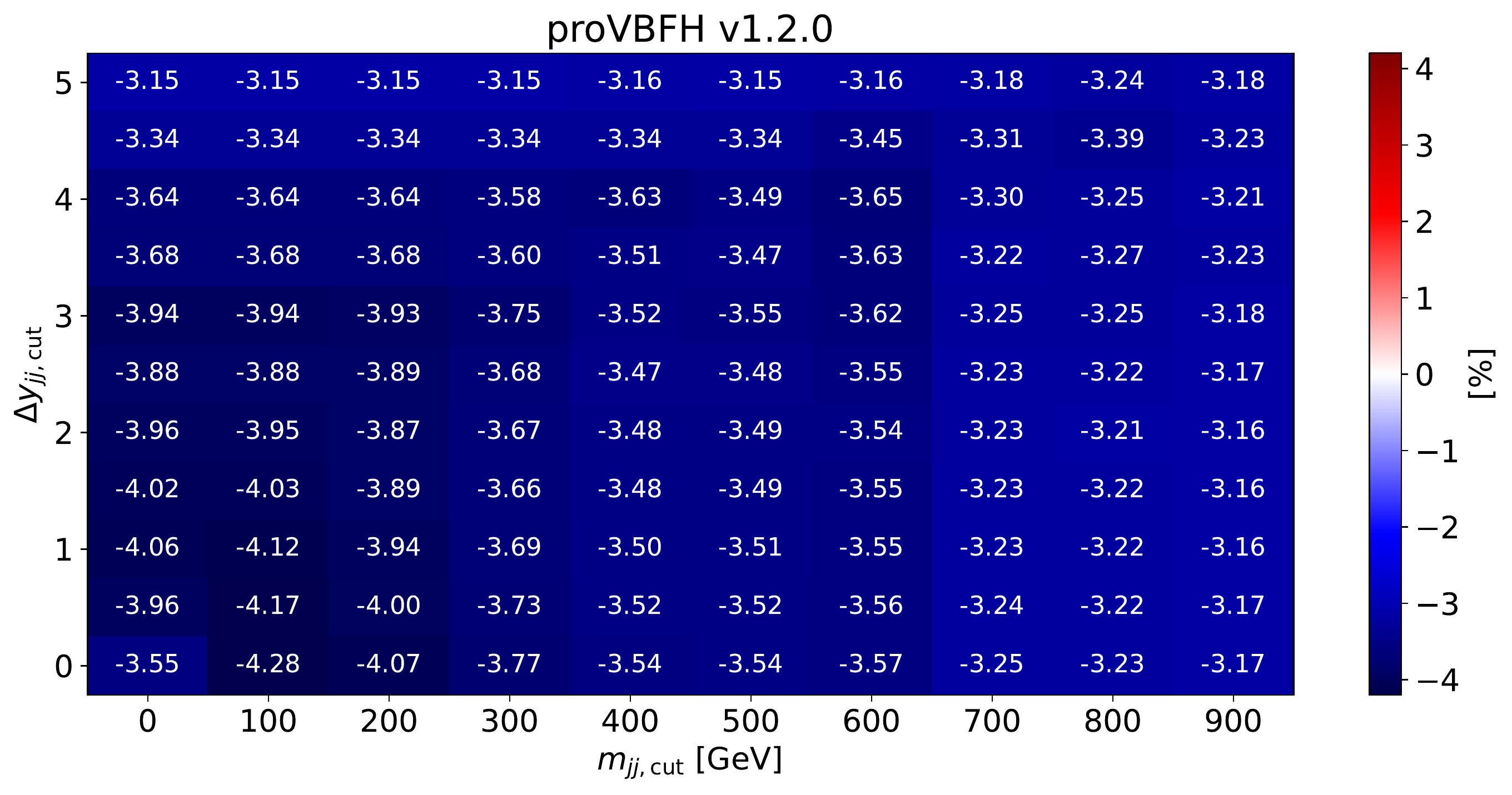}%
  \label{fig:heatmap-fact}}%
  \subfloat[Non-factorisable corrections]{\includegraphics[width=0.5\textwidth]{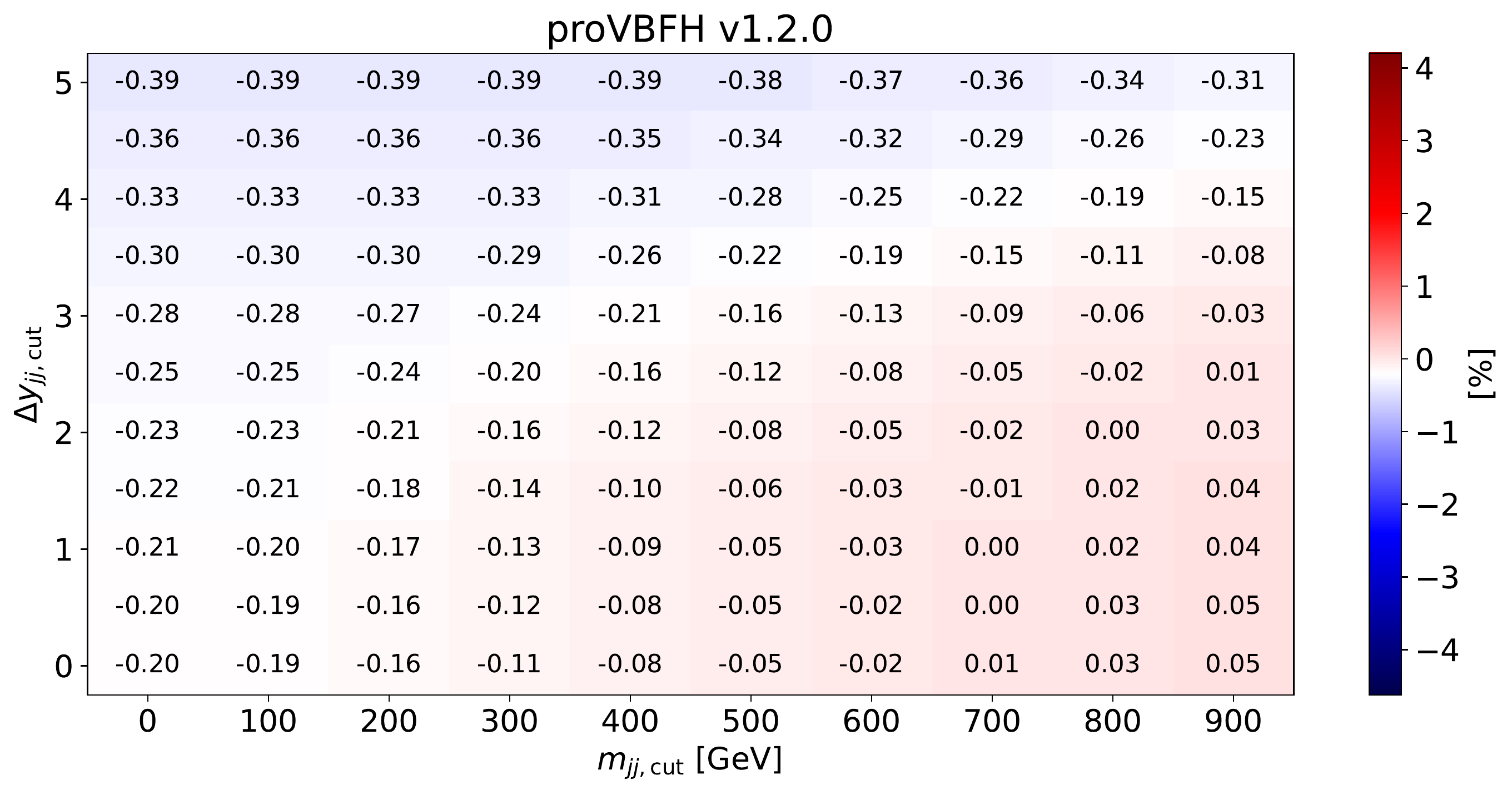}%
  \label{fig:heatmap-nonfact}}
\caption{Single Higgs VBF production: Ratio of the factorisable (a)
  and non-factorisable (b) NNLO corrections relative to LO for
  fiducial cross sections with two $R=0.4$ anti-$k_t$ jets satisfying
  $p_t > 25\GeV$ and $|y_j|<4.5$, as a function of the $m_{jj}$ and
  $\Delta y_{jj}$ selection cut.}
  \label{fig:heatmap}
\end{figure*}
\begin{figure*}[htb!]
  \centering
  \subfloat[Factorisable corrections]{\includegraphics[width=0.5\textwidth]{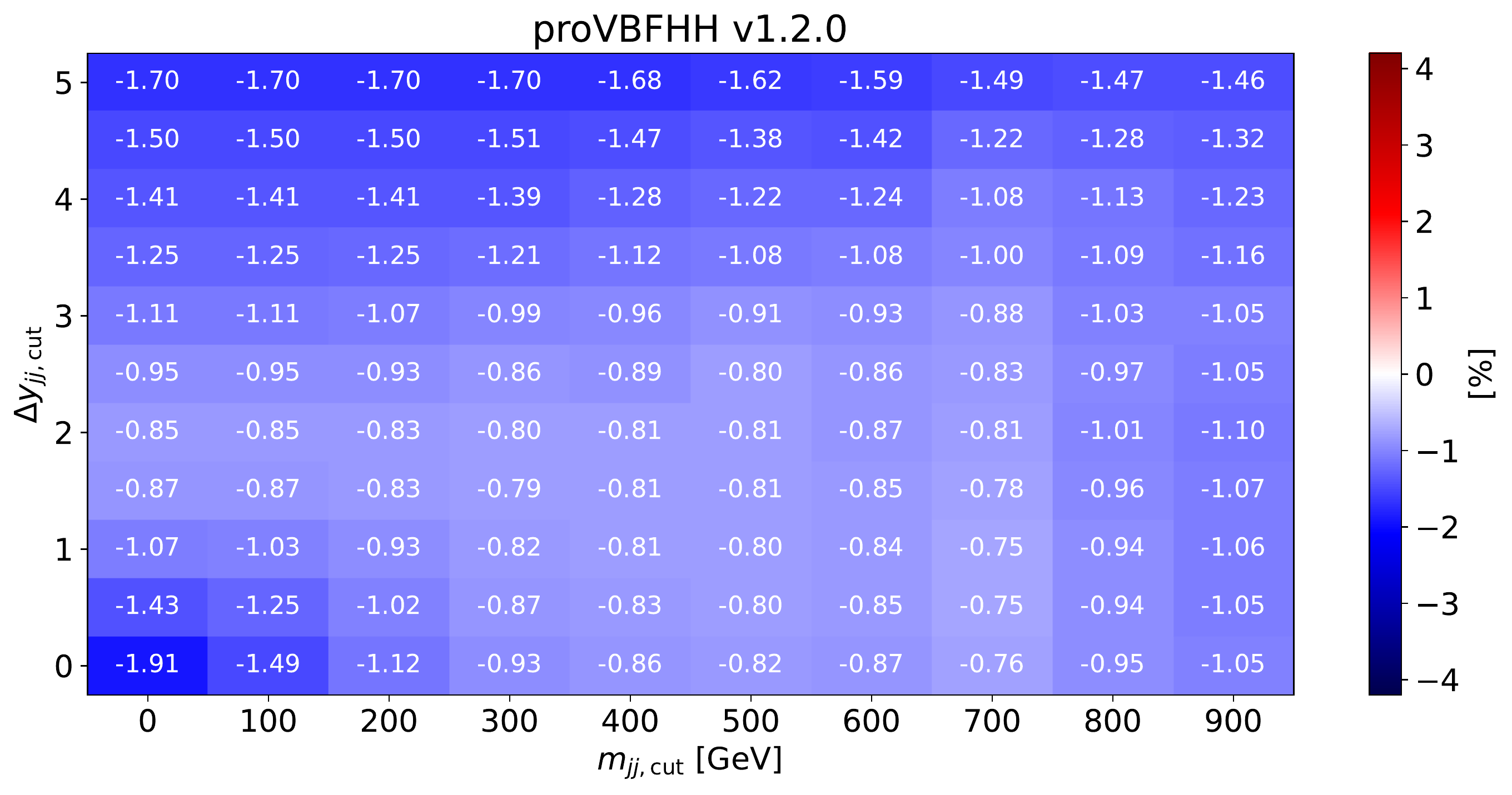}%
  \label{fig:heatmap-fact-hh}}%
  \subfloat[Non-factorisable corrections]{\includegraphics[width=0.5\textwidth]{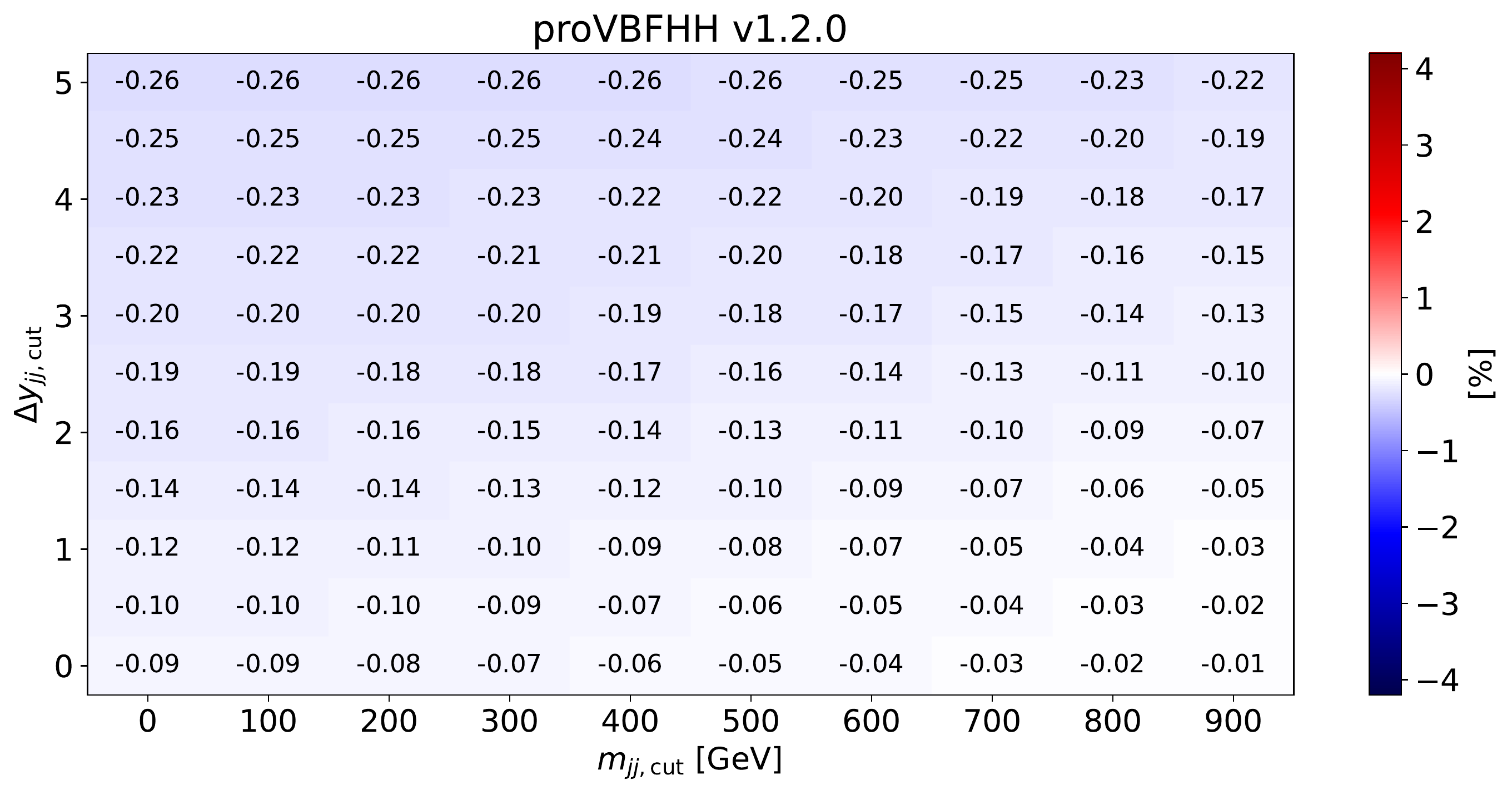}%
  \label{fig:heatmap-nonfact-hh}}
\caption{
Double Higgs VBF production: Ratio of the factorisable (a)
  and non-factorisable (b) NNLO corrections relative to LO for
  fiducial cross sections with two $R=0.4$ anti-$k_t$ jets satisfying
  $p_t > 25\GeV$ and $|y_j|<4.5$, as a function of the $m_{jj}$ and
  $\Delta y_{jj}$ selection cut.}
  \label{fig:heatmap-hh}
\end{figure*}

Following the discovery of the Higgs boson in
2012~\cite{Aad:2012tfa,Chatrchyan:2012xdj}, it has become a primary
focus of the experimental program of the Large Hadron Collider (LHC)
to measure its properties and in particular its couplings to itself
and to the other Standard Model particles~\cite{DiMicco:2019ngk}. One
of the key channels for studying the Higgs boson is the Vector-Boson
Fusion (VBF) production mode, where the Higgs boson is produced
together with two (typically) hard and forward jets.
This process has been the focus of several recent fixed order
theoretical
calculations~\cite{Figy:2003nv,Bolzoni:2010xr,Jager:2014vna,Cacciari:2015jma,Cruz-Martinez:2018rod,Dreyer:2016oyx,Campanario:2018ppz}.

A common point between all these calculations is that they are
performed in the factorised approximation, which corresponds to the
limit where partons from the two colliding protons are treated as
coming from two identical copies of QCD that interact exclusively
through the electroweak sector.
When all emissions are integrated over, this approximation is
referred to as the structure function approach~\cite{Han:1992hr}.
Due to colour conservation, this approach is exact up to NLO, but
starts to be violated from NNLO onwards, where
colour-singlet two-gluon exchanges between the incoming partons are
neglected.
Since these non-factorisable contributions are colour suppressed
compared to their factorisable counterparts, it has generally been
assumed that they can also safely be neglected~\cite{Bolzoni:2010xr}.
Recently it has been shown that the impact of the non-factorisable corrections at NNLO
 can be estimated
in the so-called eikonal approximation~\cite{Liu:2019tuy}.
Although this calculation confirms that their impact is moderate, it
was found that these contributions also receive a $\pi^2$-enhancement
due to their connection to the Glauber scattering phase, which
partially overcomes the effects of colour suppression.

Given these findings, the purpose of this paper is two-fold.
Firstly, we investigate the validity of the approximation employed in
Ref.~\cite{Liu:2019tuy} for single Higgs production outside of tight
VBF cuts, in order to estimate the leading non-factorisable
corrections on the inclusive VBF cross section.  We then conduct an
in-depth phenomenological study of the factorisable and
non-factorisable corrections, and establish the relative impact of the
latter for a range of selection cuts and observables.  Secondly, we
extend the calculation of Ref.~\cite{Liu:2019tuy} to study the impact
of non-factorisable corrections to the production of a pair of Higgs
bosons in VBF. In this case, contrary to single Higgs, it is well
known that the rather small LO cross-section is the result of delicate
cancellations of more than one order of magnitude between the
different Feynman diagrams that contribute to the process, shown in
figure~\ref{fig:feyn-VBFHH}.  While QCD radiative corrections in the
factorisable approximation affect equally all Born diagrams and are
not expected to spoil this cancellation, the same cannot be expected a
priori for the non-factorisable ones.  As we will demonstrate
in this paper, unitarity ensures that non-factorisable radiative corrections
preserve the delicate pattern of cancellations observed at leading order,
leading to overall suppressed contributions, both at the inclusive and at the
differential level.

In figure~\ref{fig:heatmap}, we provide a summary of the impact of
$\mathcal{O}(\as^2)$ corrections to single Higgs VBF production as a
function of the selection cuts on the rapidity separation
$\Delta y_{jj}$ and the invariant mass $m_{jj}$.
Figure~\ref{fig:heatmap-fact} shows the ratio of the factorisable
corrections to the LO cross section. The corrections have only a mild
dependence on the cuts, decreasing from roughly $-4\%$ at low cuts to
around $-3\%$ at larger cut values. The non-factorisable corrections
shown in figure~\ref{fig:heatmap-nonfact} on the other hand show a
stronger dependence on the cuts. They increase in size with an
increase in the $\Delta y_{jj}$ cut, and decrease as the $m_{jj}$ cut
increases until they become positive but still small at very large
$m_{jj}$ cut values. In general they are suppressed by an order of
magnitude compared to the factorisable corrections.
In figure~\ref{fig:heatmap-hh} we show the same comparison but for
di-Higgs production.
As can be seen in figure~\ref{fig:heatmap-fact-hh} the factorisable
corrections have a more complicated dependence on the cut values
compared to single Higgs VBF production, first decreasing with an
increase in both cuts and then finally increasing in size as both cuts
become large.
However, the non-factorisable corrections shown in
figure~\ref{fig:heatmap-nonfact-hh} have a similar kinematical
dependence to the single-Higgs case, although they are somewhat
suppressed as is the case for their factorisable analogue.

We note that in addition to the non-factorisable corrections studied
in this paper, a number of known perturbative corrections to VBF Higgs
production are usually neglected. These include $t$/$u$-channel
interference and $s$-channel contributions~\cite{Ciccolini:2007ec},
single-quark line contributions~\cite{Harlander:2008xn}, and loop
induced interferences between VBF and gluon-fusion Higgs
production~\cite{Andersen:2007mp}. These corrections are small within
typical VBF cuts and we do not consider them here. The NLO corrections
in the electroweak coupling have also been studied in
Ref.~\cite{Ciccolini:2007ec}.

The rest of the paper is structured as follows:
in section~\ref{sec:vbfh_intro} we provide a review of the known QCD
corrections to VBF single Higgs production, and describe how to perform a similar estimate of the non-factorisable
corrections to di-Higgs production in the eikonal approximation.
In section~\ref{sec:results} we compare factorisable and
non-factorisable corrections for VBF single Higgs production in a realistic
setup.
In section~\ref{sec:dihiggs} we discuss the impact of the
non-factorisable corrections to VBF di-Higgs production.
In section~\ref{sec:conclusion} we give our conclusions.

\section{QCD corrections in VBF Higgs production}
\label{sec:vbfh_intro}
\subsection{Factorisable corrections}
\label{sec:nnlo_fact}

\begin{figure}[h!]
  \centering
  \subfloat[]{\includegraphics[width=0.45\linewidth]{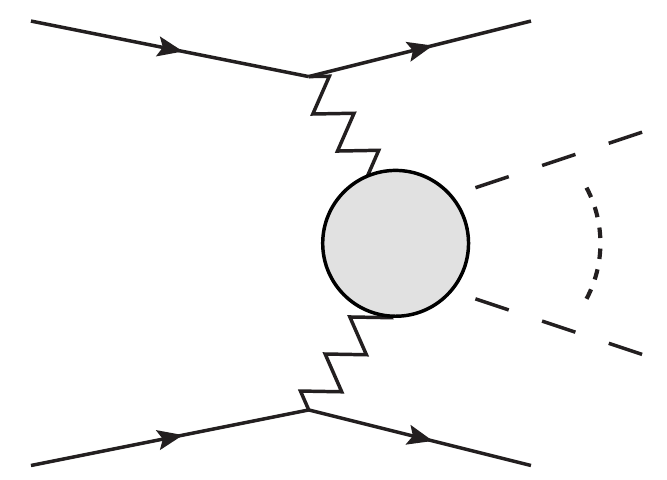}%
  \label{fig:nf-Born}}%
  \subfloat[]{\raisebox{0.0\height}{\includegraphics[width=0.45\linewidth]{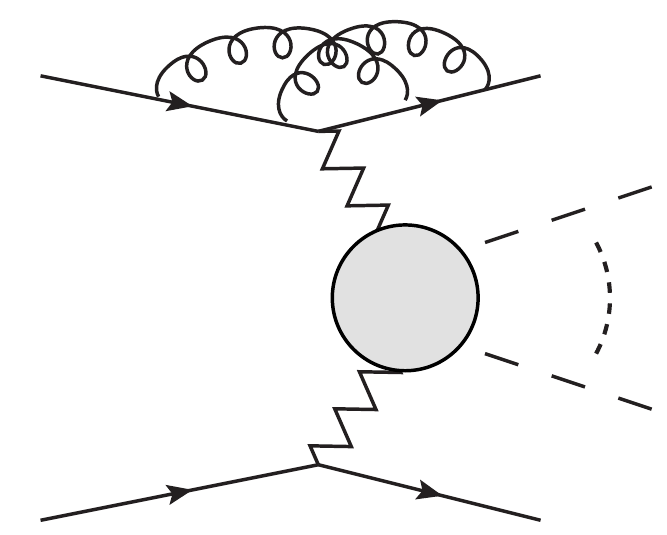}}%
  \label{fig:fact}}
  \caption{Born diagram for the production of $n$ Higgs bosons in VBF (a) and representative 2-loop factorisable corrections (b).}
\end{figure}

Both in single and double Higgs production via VBF, the Higgs bosons
are emitted by the electroweak vector bosons exchanged between the two
scattering partons.  Schematically, the Born process for the emission
of an arbitrary number of Higgs bosons can be depicted as in
figure~\ref{fig:nf-Born}.

In the factorised approximation, the VBF  cross
section 
is then expressed as a double deep inelastic scattering (DIS)
process, see Fig.~\ref{fig:fact}, 
for which the cross section is given by~\cite{Han:1992hr}
\begin{align}
  \label{eq:vbfh-dsigma}
  d\sigma = &\,\sum_V\frac{4\sqrt{2}G_F^3 m_V^8}{s}
  \Delta_V^2(Q_1^2)
  \Delta_V^2(Q_2^2) \,
  d\Omega_\text{VBF} \notag
  \\
  &\times
    \mathcal{W}^V_{\mu\nu}(x_1,Q_1^2)
    \mathcal{M}^{V,\mu\rho}
    \mathcal{M}^{V*,\nu\sigma}
  \mathcal{W}^{V}_{\rho\sigma}(x_2,Q_2^2)\,.
\end{align}

Here $V= W^\pm, Z$ corresponds to the mediating boson with mass $m_V$
and squared propagator $\Delta_V^2$, $G_F$ is Fermi's constant,
$\sqrt{s}$ is the collider centre-of-mass energy, $Q_i^2=-q_i^2$ and
$x_i = Q_i^2/(2 P_i\cdot q_i)$ are the usual DIS variables,
$\mathcal{W}^V_{\mu\nu}$ is the hadronic tensor and $d\Omega_\text{VBF}$
is the VBF phase space. The matrix element of the vector-boson
fusion sub-process is denoted as $\mathcal{M}^{V,\mu\nu}$. 

The hadronic tensor can be expressed as
\begin{multline}
  \label{eq:hadr-tensor}
  \mathcal{W}^V_{\mu\nu}(x_i,Q_i^2) = 
  \Big(-g_{\mu\nu}+\frac{q_{i,\mu}q_{i,\nu}}{q_i^2}\Big) F_1^V(x_i,Q_i^2)
  \\
  + \frac{\hat{P}_{i,\mu}\hat{P}_{i,\nu}}{P_i\cdot q_i} F_2^V(x_i,Q_i^2)
  + i\epsilon_{\mu\nu\rho\sigma}\frac{P_i^\rho q_i^\sigma}{2 P_i\cdot q_i} 
  F_3^V(x_i,Q_i^2)\,,
\end{multline}
where we have defined
$\hat{P}_{i,\mu} = P_{i,\mu} - \tfrac{P_i \cdot q_i}{q_i^2} q_{i,\mu}$
and $F^V_i(x,Q^2)$ are the standard DIS structure functions with
$i=1,2,3$.
\vspace{0.5cm}

\begin{figure}[h!]
  \centering
  \includegraphics[width=0.45\linewidth]{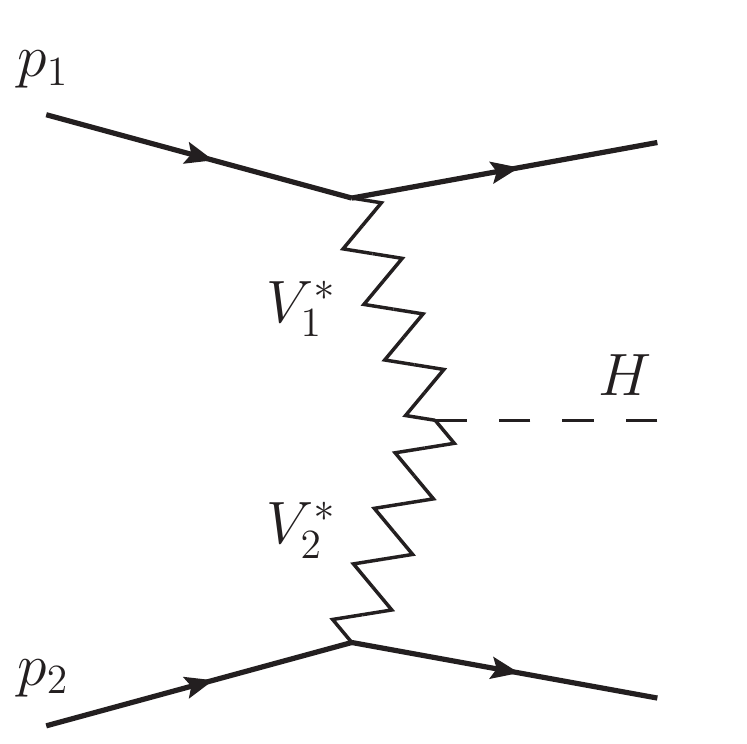}%
  \caption{Born diagram $T$ for single Higgs VBF production.}
  \label{fig:singleH}
\end{figure}

For single Higgs production, given by the diagram $T$ in
Fig.~\ref{fig:singleH}, $\mathcal{M}^{V,\mu\nu}$ can be written as
\begin{equation}
  \label{eq:matel}
  \mathcal{M}^{V,\mu\nu}= g^{\mu\nu}\,.
\end{equation}
By using the known DIS coefficient functions up to order
$\as^3$~\cite{Vogt:2004mw,Moch:2004xu,Vermaseren:2005qc,Moch:2007rq,Buehler:2011ev},
this can be used to evaluate the inclusive VBF cross section to single
Higgs production up to N$^3$LO in the factorised approximation.
By combining an inclusive NNLO calculation with the corresponding
fully differential NLO prediction for electroweak Higgs production in
association with three jets~\cite{Jager:2014vna}, one can obtain fully
differential results at NNLO through the projection-to-Born
method~\cite{Cacciari:2015jma} or the antenna subtraction
method~\cite{Cruz-Martinez:2018rod}.


\begin{figure*}
  \centering
  \subfloat[]{\includegraphics[width=0.25\linewidth]{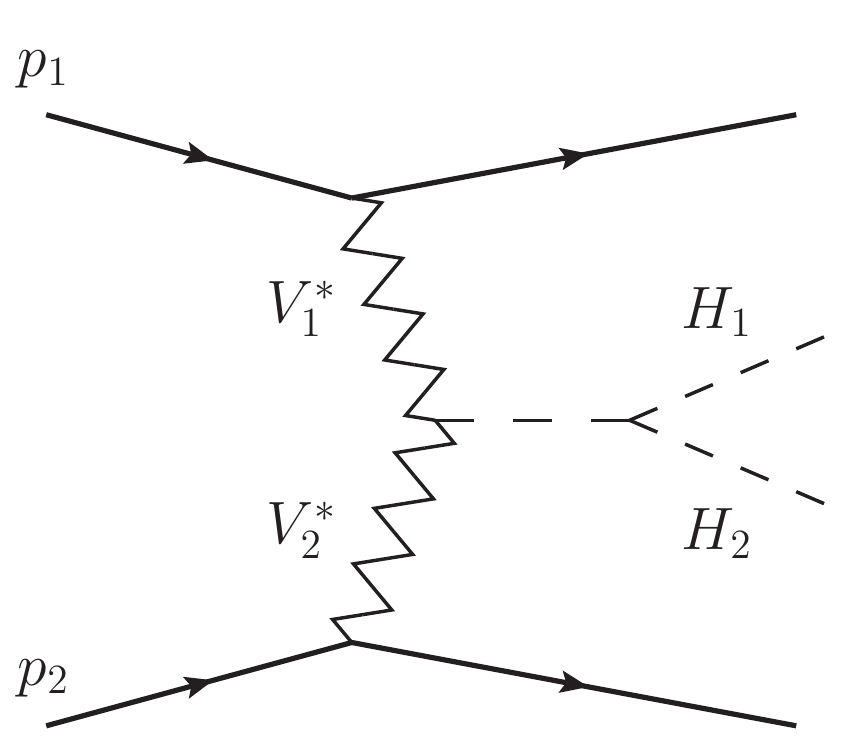}%
  \label{fig:feyn-hhh}}
  \hfill%
  \subfloat[]{\includegraphics[width=0.25\linewidth]{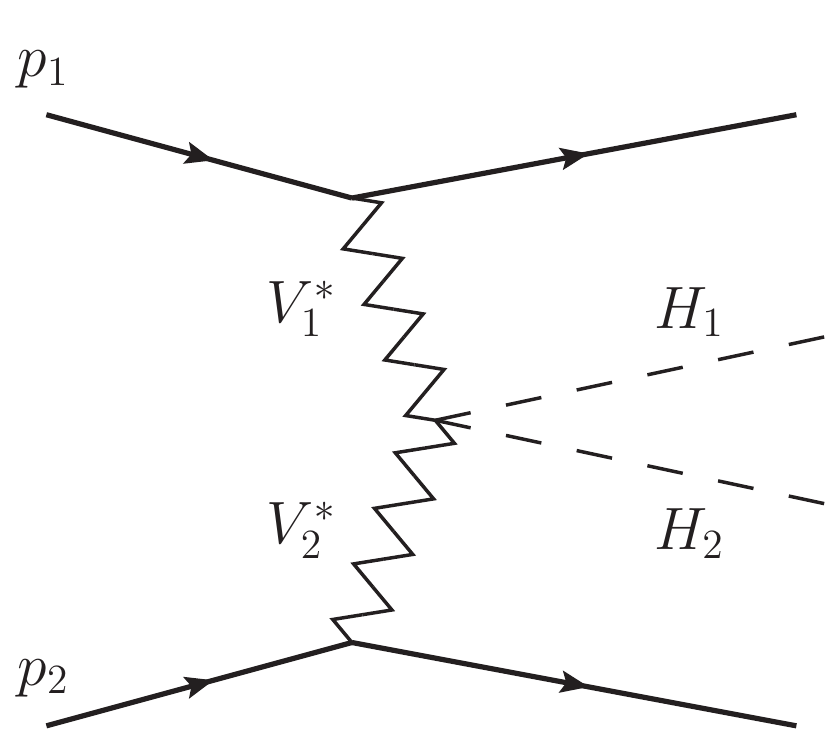}%
  \label{fig:feyn-hhVV}}
  \hfill%
  \subfloat[]{\includegraphics[width=0.25\linewidth]{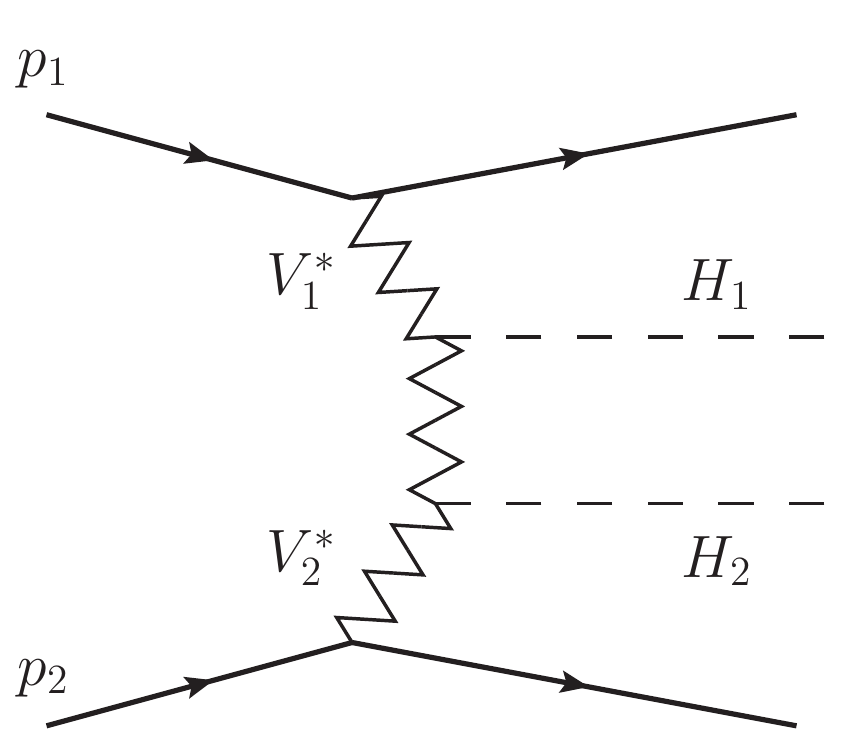}%
  \label{fig:feyn-hVV-t}}
  \hfill%
  \subfloat[]{\includegraphics[width=0.25\linewidth]{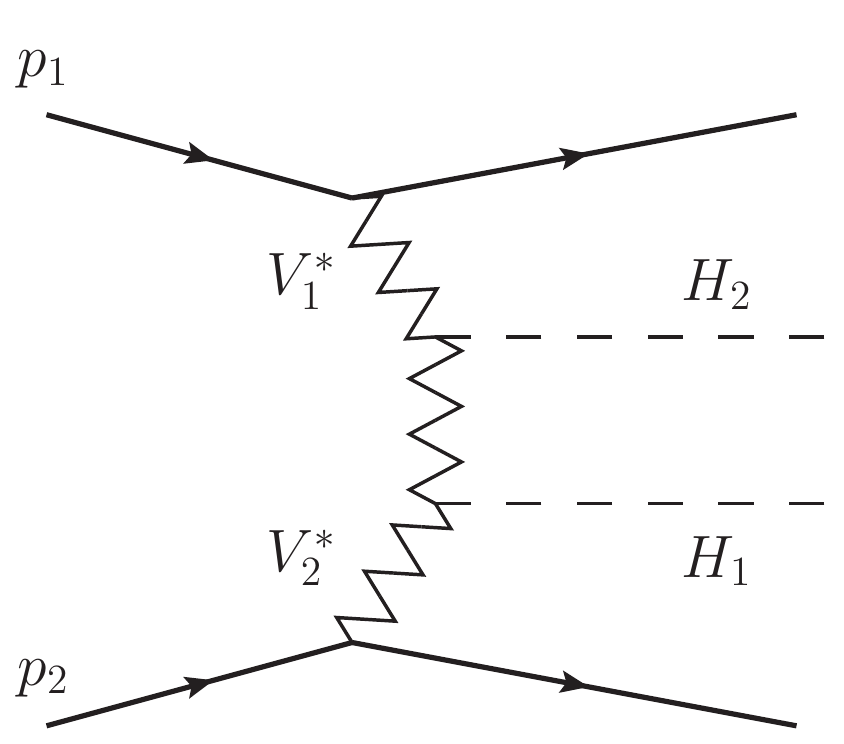}%
  \label{fig:feyn-hVV-u}}
\caption{Diagrams for Higgs pair production. (a) The $T_1$
  topology. (b) The $T_2$ topology. (c) The $B_1$ topology. (d) The
  $B_2$ topology.}
  \label{fig:feyn-VBFHH}
\end{figure*}

The factorisable QCD corrections to the di-Higgs process can be
calculated in the same way as for its single Higgs analog, expressing
the cross section in the form of eq.~(\ref{eq:vbfh-dsigma}), but
with $\mathcal{M}$ now referring to the di-Higgs matrix element.
The Higgs pair production process differs from the single Higgs case 
only at the interaction between the vector
and Higgs bosons, where an additional Higgs can arise from an
intermediate vector or Higgs boson, or from the $hhVV$ quartic
coupling.
The $VV\to hh$ sub-process at LO can be expressed as~\cite{Dobrovolskaya:1990kx}

\begin{align}
  \label{eq:VV-subproc}
  \mathcal{M}^{V,\mu\nu} &= 
  \bigg[\bigg(1
  + \,\frac{4m_V^2}{\Delta_V}\,  + \,\frac{6 \nu \lambda}{\Delta_H}
  \bigg)g^{\mu\nu}\,\\
  &+ \,\frac{m_V^2}{\Delta_V}
  \frac{(2 k_1^\mu + q_1^\mu)(k_2^\nu - k_1^\nu - q_1^\nu)}{m_V^2 - i \Gamma_V m_V}
  \bigg] + (k_1 \leftrightarrow k_2)\,, \nonumber
\end{align}
where we have defined the propagators
\begin{equation}
\begin{split}
\Delta_V = (q_1 + k_1)^2 - m_V^2 + i\Gamma_V m_V, \\
\Delta_H = (k_1 + k_2)^2 - m_H^2 + i \Gamma_H m_H 
\end{split}
\end{equation}
and $k_1, k_2$ are the momenta of the final state Higgs bosons and
$\lambda$ and $\nu$ are the trilinear Higgs self-coupling and the vacuum
expectation value of the Higgs field respectively.
The matrix element arises from the four Feynman diagrams shown in
Figure~\ref{fig:feyn-VBFHH}, which we label $T_1$, $T_2$, $B_1$ and
$B_2$. We stress here that, while we are including the bosons' widths for completeness,
they play no role for the estimation of QCD corrections to Higgs production in VBF.

\subsection{ Non-factorisable corrections}
\label{sec:nnlo_nonfact}
\begin{figure}[h]
  \centering
  \subfloat[]{\includegraphics[width=0.45\linewidth]{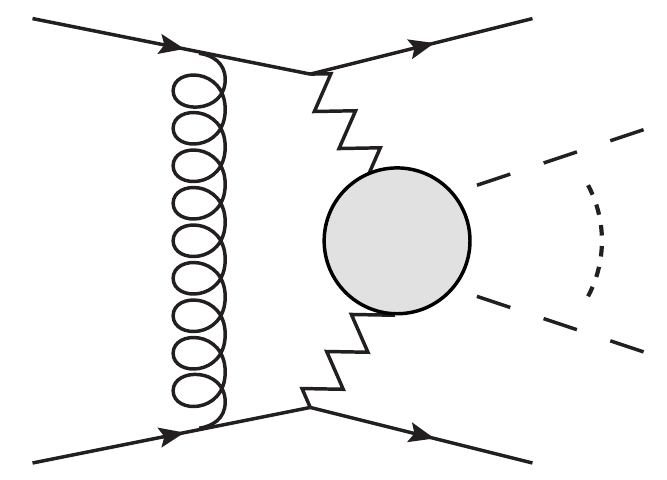}%
  \label{fig:nf-1loop}}%
  \subfloat[]{\raisebox{0.0\height}{\includegraphics[width=0.45\linewidth]{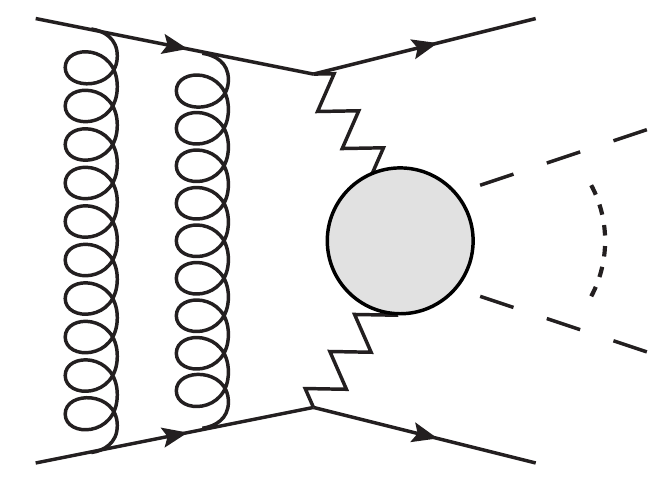}}%
  \label{fig:nf-2loop}}
  \caption{Generic form of non-factorisable 1-loop (a) and 2-loop (b) corrections to the production
  of $n$ Higgs boson.}
  \label{fig:non-fact}
\end{figure}

The factorisable approach described above, which includes diagrams
such as the one represented in figure~\ref{fig:fact}, is exact up
to NLO due to colour conservation.
At NNLO this is no longer true, as in particular two gluons in a colour singlet
state can be emitted between the two quark lines, as shown in
figure~\ref{fig:non-fact}.
As the gluons have to be in a colour singlet state, these diagrams will be
 colour suppressed compared to their factorisable counterparts. 
For this reason it has long been argued that they can be neglected
when considering NNLO corrections to VBF~\cite{Bolzoni:2010xr}.

Due to the complexity involved in computing the two-loop
non-factorisable corrections, very little has been known about them
beyond the fact that they are colour suppressed.
However, very recently~\cite{Liu:2019tuy} significant progress was
made, when it was shown that the corrections can be estimated within
the eikonal
approximation~\cite{Cheng:1969ab,Chang:1969by,Cheng:1970jk,Lipatov:1976zz}.
This calculation exploits the fact that when typical VBF cuts are
applied, the VBF cross section can be expanded in the ratio of the
leading jet transverse momentum over the total partonic centre-of-mass
energy
\begin{align}
  \xi = \frac{p_{t,j_1}}{\sqrt{s}}. \label{eq:defxi}
\end{align}
In this kinematical configuration, the authors of Ref.~\cite{Liu:2019tuy} conclude that
the non-factorisable corrections receive a $\pi^2$-enhancement
connected to the presence of a Glauber phase, which can partially compensate their
colour suppression.
Indeed, it turns out that for VBF single Higgs production, 
the non-factorisable corrections can contribute up to $1\%$  
in certain regions of phase space, making them larger than the
factorisable N$^3$LO corrections. In what follows we will use the same approximation
to estimate the impact of non-factorisable corrections for the case of double Higgs production as well.

%%%%%% CHANGE HERE
In order to see how the NNLO non-factorisable corrections can be estimated in the eikonal 
approximation both for single and double Higgs production, 
let us consider a generic VBF Born diagram, which we will call $D$, for the production of an in principle 
arbitrary number of
Higgs bosons, see Fig.~\ref{fig:nf-Born}.
In what follows this diagram will represent either the Born diagram for VBF single Higgs production $T$ of Fig.~\ref{fig:singleH}, 
or any of the Born diagrams  for double Higgs production
$T_1$, $T_2$, $B_1$ or $B_2$ 
in Fig.~\ref{fig:feyn-VBFHH}.

It is important to stress here that, somewhat counterintuitively, 
we will be considering QCD corrections on each single diagram separately, and not on the full Born matrix element.
Since we are interested in computing the NNLO QCD corrections to this class of processes, 
we imagine dressing the diagram $D$ with 1-loop or 2-loop QCD corrections,
as depicted in Fig.~\ref{fig:non-fact}, 
where we provide two representative diagrams for illustration only.

It turns out that, at least up to two loops in QCD, 
we can limit ourselves to diagrams where the gluons are
in a colour-singlet configuration, i.e.\ 
exchanged between the two quark lines. All other configurations do not contribute to the cross-section 
due to colour conservation. 
Therefore, the calculation of the one- and two-loop QCD corrections in the eikonal approximation reduces effectively 
to the corresponding calculation in QED, with the colour-averaged effective coupling
\begin{equation}
\widetilde{\alpha}_s = \left( \frac{N_c^2-1}{4 N_c^2} \right)^{1/2} \alpha_s\,.
\label{eq:effcoupl}
\end{equation}
Following Ref.~\cite{Liu:2019tuy}, 
let us consider the process
\begin{equation}
q(p_1) + q(p_2) \to q(p_3) + q(p_4) + X(P)
\end{equation}
where $X(P)$ can represent one or multiple Higgs bosons produced in vector-boson fusion.
At leading order, we call the momenta flowing in the two vector bosons respectively
\begin{equation}
q_1 = p_1-p_3\,, \quad q_2 = p_2-p_4\,. \label{eq:Vmom}
\end{equation}

The leading term in the eikonal approximation can then easily be
obtained by employing light-cone coordinates, which make transparent
the separation between the dynamics in the plane spanned by the
momenta of the incoming quarks and the plane transverse to
them~\cite{Cheng:1969ab,Chang:1969by,Cheng:1970jk,Lipatov:1976zz}. For a momentum
$k^\mu$ we indicate by $k^{\pm}$ the light-cone coordinates and by
$\bm{k}$ those in the transverse plane, i.e.\ we write
\begin{equation}
k^\mu = (k^+, k^-, \bm{k})\,, \quad k^\pm = \frac{k^0 \pm k^3}{\sqrt{2}}\,, \quad \bm{k} = (k_1, k_2)\,,
\end{equation}
and we choose a reference frame such 
that the incoming quark momenta have each one light-cone component
different from zero
$$p_1^\mu = (0,p_1^{-}, \bm{0})\,, \qquad p_2^\mu = (p_2^{+},0, \bm{0})\,.$$

\begin{figure}[h]
  \centering
  \subfloat[]{\includegraphics[width=0.45\linewidth]{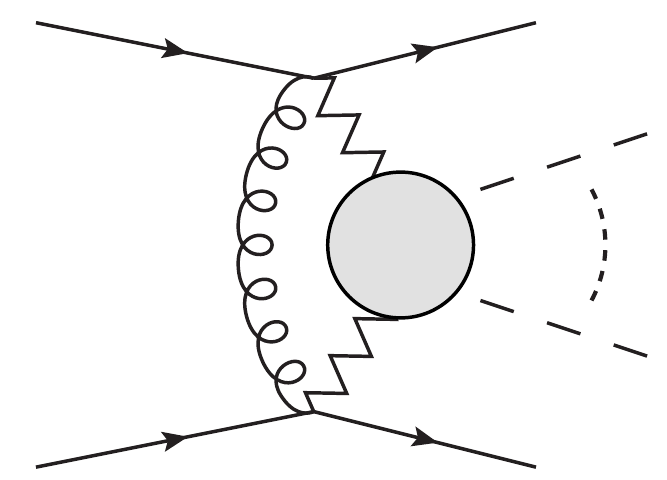}%
  \label{fig:nf-1loop_transv}} %
  \subfloat[]{\raisebox{0.0\height}{\includegraphics[width=0.45\linewidth]{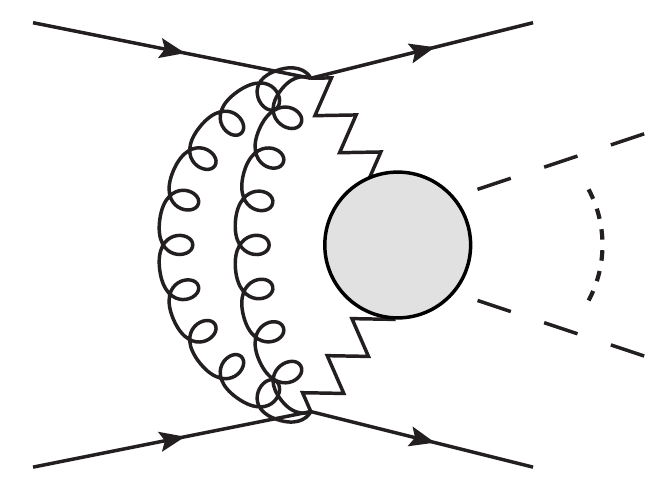}}%
  \label{fig:nf-2loop_transv}}
  \caption{Non-factorisable 1-loop (a) and 2-loop (b) corrections in the eikonal approximation. Notice that
  these are two-dimensional euclidean diagrams in the plane transverse to the incoming quark momenta.}
  \label{fig:non-fact_transv}
\end{figure}

It turns out that both at one and two loops, \emph{at leading order in the eikonal approximation},
the quark propagators coupled to the soft
gluons simplify and, after summing over all permutations of the gluons and the vector bosons, 
the quark propagators recombine in terms of delta functions of the light-cone components of the 
loop momenta. 
This allows one to effectively decouple the light-cone dynamics from the  one in the two-dimensional 
plane transverse to the momenta of the incoming quarks
and  one is left with the  calculation of the effective \emph{two-dimensional} loop diagrams shown
schematically in Fig.~\ref{fig:non-fact_transv}.  
Extra care has to be taken when considering diagrams of type $(c)$ and $(d)$ in Fig.~\ref{fig:feyn-VBFHH}, 
where even in the eikonal approximation the light-cone components cannot be neglected in 
the propagator of the central vector boson. Their effect can nevertheless be effectively included by modifying
the mass of the central vector boson to an effective mass $M_V \to \mu_V$, whose value depends
on the light-cone components of the scattering quarks and of the Higgs boson.\footnote{We are grateful to 
A. Penin for elucidating this point to us.} We provide its explicit value
later on, see Eq.~\eqref{eq:muV}.

With this, one can easily write the one- and two-loop QCD corrections in the eikonal approximation 
in a rather compact form. 

By calling $\vbone$ and $\vbtwo$ the transverse components of the momenta $q_1$ and $q_2$ in~\eqref{eq:Vmom},
and indicating schematically with $\{\bm{q},\mu_V\}$ the set of transverse momenta of the Higgs 
bosons produced and the effective mass of the intermediate vector boson, as described above, 
we can write for the generic Born diagram $D$
\begin{align}
&\mathcal{M}_D^{(1)} = + i \widetilde{\alpha}_s \chi_D^{(1)}(\vbone, \vbtwo, \{\bm{q},\mu_V\})\, \mathcal{M}_D^{(0)}\,, \\
&\mathcal{M}_D^{(2)} = - \frac{\widetilde{\alpha}_s^2 }{2!} \chi_D^{(2)}(\vbone, \vbtwo,\{\bm{q},\mu_V\})\, \mathcal{M}_D^{(0)}\,,
\end{align}
where $\mathcal{M}_D^{(n)} $ are the corrections to the Born diagram $D$ coming from the exchange of $n$ gluons,
 $\chi_D^{(n)}(\vbone, \vbtwo, \{\bm{q},\mu_V\})$  are functions which depend on the (transverse) 
 kinematics of the corresponding Born diagram and on the effective mass $\mu_V$,
 and the effective coupling $\widetilde{\alpha}_s$ was defined in eq.~\eqref{eq:effcoupl}.
 Finally, the factor $1/2!$ comes from the symmetrisation of the two identical gluons~\cite{Liu:2019tuy}.
We stress once more that, if we are interested in double Higgs production, this happens separately for each of the Born
diagrams in Fig.~\ref{fig:feyn-VBFHH}. 
We also remind the reader that this is true as long as we limit ourselves to colour-singlet gluon exchange.

Given the considerations above, it is easy to see that QCD corrections to
the Born diagram of single Higgs production $T$, or to $T_{1,2}$ for double Higgs, reduce to the computation of
a two-dimensional one- or two-loop triangle-like integral, while the corrections to $B_{1,2}$, 
involve the computation of more complicated box-like loop integrals. 
Moreover, it should  also be clear that for $T$, $T_1$ and $T_2$, the QCD corrections
only depend on the momenta $\vbone$, $\vbtwo$ and are therefore equal in all three cases. 
Putting everything together, we find 
 similarly to Ref.~\cite{Liu:2019tuy} 
\begin{align}
&\chi_{T}^{(1)}(\vbone, \vbtwo) = \frac{1}{\pi} \int \frac{d^2 \bm{k}}{\bm{k}^2 + \lambda^2} \nonumber \\
&\times  \frac{\vbone^2 + M_V^2}{(\bm{k}-\vbone)^2+M_V^2}  \frac{\vbtwo^2 + M_V^2}{(\bm{k}+\vbtwo)^2+M_V^2} \\
&\chi_{T}^{(2)}(\vbone, \vbtwo) = \frac{1}{\pi^2} \int \left( \prod_{i=1}^2 \frac{d^2 \bm{k_i}}{\bm{k_i}^2 + \lambda^2} \right) \nonumber \\
&\times  \frac{\vbone^2 + M_V^2}{(\bm{k}_{12}-\vbone)^2+M_V^2}  \frac{\vbtwo^2 + M_V^2}{(\bm{k}_{12}+\vbtwo)^2+M_V^2}\,,
\end{align}
where we defined $\bm{k}_{12} = \bm{k}_1 + \bm{k}_2$\ and have introduced a fictitious gluon mass $\lambda$
to regulate the residual IR divergences. Also, we have removed the dependence on the momenta $\{\bm{q},\mu_V \}$ 
since for these diagrams 
$\vbone+\vbtwo = \bm{q}$ and there is no vector boson propagator that depends on $\mu_V$.
\newline

Let us consider now the two box-like topologies, which have a non-trivial dependence on the momenta of the two Higgs bosons.
Calling their momenta  $q_3$ and $q_4$ and using 
$\bm{q}_4 = -\bm{q}_1-\bm{q}_2-\bm{q}_3$ (all momenta are incoming), we find

\begin{align}
  \label{eq:chi-HH}
  &\chi_{B_1}^{(1)}(\vbone, \vbtwo, \bm{q}_3,\mu_V) = \frac{1}{\pi} \int \frac{d^2 \bm{k}}{\bm{k}^2 + \lambda^2} \nonumber \\
  &\times \frac{\vbone^2 + M_V^2}{(\bm{k}-\vbone)^2+M_V^2}  \frac{\vbtwo^2 + M_V^2}{(\bm{k}+\vbtwo)^2+M_V^2}
 \frac{t + \mu_V^2}{(\bm{k}-\bm{q}_{13})^2+\mu_V^2} \nonumber \\ & \nonumber \\
 %%%%%
  & \chi_{B_1}^{(2)}(\vbone, \vbtwo,\bm{q}_3,\mu_V) = \frac{1}{\pi^2} \int \left( \prod_{i=1}^2 \frac{d^2 \bm{k_i}}{\bm{k_i}^2 + \lambda^2} \right) \nonumber \\
&\times  \frac{\vbone^2 + M_V^2}{(\bm{k}_{12}-\vbone)^2+M_V^2}  \frac{\vbtwo^2 + M_V^2}{(\bm{k}_{12}+\vbtwo)^2+M_V^2}
 \frac{t + \mu_V^2}{(\bm{k}_{12}-\bm{q}_{13})^2+\mu_V^2} \nonumber \\
%%%%%%                                                                         
  &\nonumber \\ 
  &\chi_{B_2}^{(j)}(\vbone, \vbtwo,\bm{q}_4,\mu_V) = \chi_{B_1}^{(j)}(\vbone, \vbtwo, \bm{q}_3,\mu_V)\Big|_{ \begin{smallmatrix}
q^\mu_3 \leftrightarrow q^\mu_4 \\ t \leftrightarrow u
\end{smallmatrix} }\,, \quad j=1,2\,.
\end{align}
where we put $\bm{q}_{ij} = \bm{q}_i + \bm{q}_j$ and 
 defined in addition the ``transverse-plane'' Mandelstam variables $s= (\vbone+\bm{q}_1)^2$\,, $t= (\vbone+\bm{q}_3)^2$\,,
$u= (\vbone+\bm{q}_4)^2$\,, as well as
\begin{align}
  \mu_V^2 &= M_V^2 - M_H^2 - \bm{q}^2_{3} - 2 q_{3}^+ q_1^- \nonumber \\
  & = M_V^2 - M_H^2 - \bm{q}^2_{4} - 2 q_{4}^- q_2^+\,. \label{eq:muV}
\end{align}
Similarly to the previous case, we regulated the residual IR divergences
with a gluon mass $\lambda$. Notice that in defining $\chi_{B_2}$
we need to swap the entire four-momenta $q_3^\mu \leftrightarrow q_4^\mu$,
which also affects the value of $\mu_V$ as defined in Eq.~\eqref{eq:muV}.
  
The integrals above can be computed in many different ways, most
notably making use of the reduction of all 1-loop and certain 2-loop
$n$-point functions, with $n \geq3$, in $d=2$ space-time dimension, to
lower-point topologies.
Also a direct computation of the integrals using their Feynman
parameter representation can be attempted, which turns out to be
particularly simple for the triangle integrals
$\chi_{T}^{(1)}(\vbone, \vbtwo)$ and $\chi_{T}^{(2)}(\vbone, \vbtwo)$,
see Ref.~\cite{Liu:2019tuy}.
While the analytic computation is conceptually straightforward, the
result, in particular for what concerns the box-type integrals, can
become very cumbersome due to their dependence on a large number of
scales and are not particularly illuminating.

Nevertheless, since we are dealing with two-dimensional euclidean integrals, it turns out to be entirely straightforward
to produce very compact one-fold integral representations for them by extracting the logarithmic
divergences as $\lambda \to 0$
and integrating directly 
on the 2-dimensional loop momenta in polar coordinates. This remains true at two loops,
where one can first integrate out the gluonic one-loop sub-bubble, and then proceed in the very same
way as for the one-loop integrals. This allows us to get all results as one-fold 
integrals over simple algebraic functions
and at most powers of logarithms.

We write down the results for the one- and two-loop triangles as

\begin{align}
\chi_{T}^{(1)}(\vbone, \vbtwo) &= - \ln{\left( \frac{\lambda^2}{M_V^2}\right)} 
+ f_T^{(1)} \nonumber \\
%%%%%
\chi_{T}^{(2)}(\vbone, \vbtwo) &= \ln^2{\left( \frac{\lambda^2}{M_V^2}\right)} 
- 2 \ln{\left( \frac{\lambda^2}{M_V^2}\right) } f_T^{(1)}  \nonumber \\
&+ f_T^{(2)} 
\label{eq:chitri}
\end{align}
and similarly for the boxes
\begin{align}
\chi_{B_1}^{(1)}(\vbone, \vbtwo, \bm{q}_3,\mu_V) &= - \ln{\left( \frac{\lambda^2}{M_V^2}\right)} 
+ f_B^{(1)}  \nonumber \\
%%%%%%
\chi_{B_1}^{(2)}(\vbone, \vbtwo, \bm{q}_3,\mu_V) &= \ln^2{\left( \frac{\lambda^2}{M_V^2}\right)} 
- 2 \ln{\left( \frac{\lambda^2}{M_V^2}\right) } f_B^{(1)} \nonumber 
\\ &+ f_B^{(2)}
\label{eq:chibox}
\end{align}
where the function $f_T^{(j)}$ and $f_B^{(j)}$ depend on the
corresponding transverse momenta and $\chi_{B_2}^{(j)}$ can be
obtained from $\chi_{B_1}^{(j)}$ by swapping
$q^\mu_3 \leftrightarrow q^\mu_4$ as in Eq.~\eqref{eq:chi-HH} and Eq.~\eqref{eq:muV}.
In order to write their analytic expression, we start off by
parametrising the kinematics in the two-dimensional transverse plane
as
\begin{align}
\bm{q}_1 = (q_{1x}, 0)\,, \quad \bm{q}_2 = (q_{2x}, q_{2y})\,,
\quad \bm{q}_3 = (q_{3x}, q_{3y})
\end{align}
with $q_4 = -q_1-q_2-q_3$, and we introduce the shorthand notation
$$\Delta_1 = (q_1^2 + M_V^2)\,, \quad \Delta_2 = (q_2^2 + M_V^2)
\,, \quad \Delta_t = (t + \mu_V^2)$$

The functions can then be written as follows

\begin{strip}
\vspace{0.3cm}
\begin{align}
f_T^{(1)} &= - \int_0^{2 \pi}  \frac{d\xi}{\pi} \frac{\Delta_1 \Delta_2 }{ r_{13}}
 \left( \frac{\ln( \bar{r}_1 )}{r_1 r_{12} r_{14}} + \frac{\ln( \bar{r}_3 )}{r_3 r_{23} r_{34}}   \right) 
 + \left( \begin{array}{c} r_1 \leftrightarrow r_2 \\ r_3 \leftrightarrow r_4 \end{array} \right) \label{eq:tri1} \\
 %%%%%
f_T^{(2)} &= - 2 \int_0^{2 \pi}  \frac{d\xi}{\pi} \frac{\Delta_1 \Delta_2 }{ r_{13}}
 \left( \frac{\ln^2( \bar{r}_1 )}{r_1 r_{12} r_{14}} + \frac{\ln^2( \bar{r}_3 )}{r_3 r_{23} r_{34}}   \right) 
+ \left( \begin{array}{c} r_1 \leftrightarrow r_2 \\ r_3 \leftrightarrow r_4 \end{array} \right) + \frac{4 \pi^2}{3} \label{eq:tri2} \\
%\end{align}
%\begin{align}
f_B^{(1)} &= - \int_0^{2 \pi}  \frac{d\xi}{\pi} \frac{\Delta_1 \Delta_2 \Delta_t }{ r_{13} r_{15} r_{35}}
 \left( \frac{ r_{35} \ln( \bar{r}_1 )}{r_1 r_{12} r_{14} r_{16}}
      + \frac{ r_{15} \ln( \bar{r}_3 )}{r_3 r_{23} r_{34} r_{36} }   
      + \frac{ r_{13} \ln( \bar{r}_5 )}{r_5 r_{25} r_{45} r_{56} }   \right) 
 + \left( \begin{array}{c} r_1 \leftrightarrow r_2 \\ r_3 \leftrightarrow r_4 \\ r_5 \leftrightarrow r_6 \end{array} \right) \label{eq:box1}\\
 %%%%%
f_B^{(2)} &= - 2 \int_0^{2 \pi}  \frac{d\xi}{\pi} \frac{\Delta_1 \Delta_2 \Delta_t }{ r_{13} r_{15} r_{35}}
 \left( \frac{ r_{35} \ln^2( \bar{r}_1 )}{r_1 r_{12} r_{14} r_{16}} 
      + \frac{ r_{15} \ln^2( \bar{r}_3 )}{r_3 r_{23} r_{34} r_{36}}
      + \frac{ r_{13} \ln^2( \bar{r}_5 )}{r_5 r_{25} r_{45} r_{56} }    \right) 
+ \left( \begin{array}{c} r_1 \leftrightarrow r_2 \\ r_3 \leftrightarrow r_4 \\ r_5 \leftrightarrow r_6  \end{array} \right) + \frac{4 \pi^2}{3} 
\label{eq:box2}
\end{align}
\end{strip}
with $r_{ij} = r_{i} - r_{j}$, $\bar{r}_j = -r_j/M_V^2$ and the six roots read
\begin{align}
r_1 &= q_{1x} \cos{\xi} - \frac{i R_1}{\sqrt{2}}\,, \quad r_2 = r_1^*\,, \nonumber \\
r_3 &= - (q_{2x} \cos{\xi} + q_{2y} \sin{\xi} ) - \frac{i R_2}{\sqrt{2}} \,, \quad r_4 = r_3^*\,, \nonumber \\
r_5 &=  ((q_{1x}+q_{3x}) \cos{\xi} + q_{3y} \sin{\xi} ) - \frac{i R_3}{\sqrt{2}} \,,
\quad r_6 = r_5^*\,,
\end{align}
where $r_j^*$ indicates complex conjugation and
\begin{align}
R_1 &= \sqrt{2M_V^2 + q_{1}^2(1-\cos{2\xi})} \nonumber \\
R_2 &= \sqrt{2(M_V^2 + q_2^2) - 2 (q_{2x}\cos{\xi} + q_{2y} \sin{\xi})^2} \nonumber \\
R_3 &= \sqrt{2(\mu_V^2 + t) - 2 ( (q_{1x}+q_{3x})\cos{\xi} + q_{3y} \sin{\xi})^2}\,.
\end{align}

The complex arguments and the three square-roots above might seem
somewhat unappealing, in particular because lengthy but fully analytic
representation can be obtained for all these functions in terms of
polylogarithms.
Nevertheless, our results involve only integrals of logarithms and
exhibit a very high degree of symmetry, both moving from one to two
loops and going from 3- to 4-point functions. Moreover it is
straightforward to rewrite the integrals to make them explicitly real,
at the price of introducing inverse trigonometric functions.  Finally,
as a curiosity, it turns out that performing the calculation in this
way the results can be effortlessly generalised to higher-point
integrals, i.e.\ for an arbitrary number of Higgs bosons in the final
state.

With the definitions above, the non-factorisable QCD corrections to the 
total amplitude for single and double Higgs production can be written, 
respectively, as
\begin{align}
\mathcal{M}_{H} = \sum_j \mathcal{M}_{H}^{(j)}\,, \qquad
\mathcal{M}_{HH} = \sum_j \mathcal{M}_{HH}^{(j)} \,,
\end{align}
where for single Higgs we have simply
\begin{align}
\mathcal{M}_H^{(j)} = \mathcal{M}_T^{(j)}\,,
\end{align}
while for double Higgs we find
\begin{align}
\mathcal{M}_{HH}^{(j)} = \mathcal{M}_{T_1}^{(j)}+\mathcal{M}_{T_2}^{(j)}+\mathcal{M}_{B_1}^{(j)}+\mathcal{M}_{B_2}^{(j)}\,,
\label{eq:amplHH}
\end{align}
which of course implies a much richer interference pattern. More explicitly, we find for the cross-section for
single Higgs production
\begin{align}
  \mathrm{d}\sigma^{\mathrm{NNLO}}_{H,\mathrm{nf}} =
\widetilde{\alpha}_s^2\, \chi^H_{\mathrm{nf}}(\vbone,\vbtwo) \; \mathrm{d}\sigma^{\mathrm{LO}}
  \label{eq:nnlo-nf_H}
\end{align}
where $\mathrm{d}\sigma^{\mathrm{LO}}$ is the leading-order cross section given in
\eqref{eq:vbfh-dsigma}, $\widetilde{\alpha}_s$ is the effective coupling in eq.~\eqref{eq:effcoupl}, 
and the NNLO non-factorisable contributions only depend on the functions $f_T^{(j)}$ through
\begin{align}
  \chi_{\mathrm{nf}}^H (\vbone,\vbtwo) &= \left[\chi_T^{(1)}(\vbone,\vbtwo)\right]^2 - \chi_T^{(2)}(\vbone,\vbtwo)\nonumber \\
&= \left[f_T^{(1)}\right]^2 - f_T^{(2)}\,.
\end{align}
As an illustration, and in order to compare this case to di-Higgs production, 
it is useful to compute the corrections in the limit where all transverse scales become small compared to the
vector-boson mass, i.e.\ $\bm{q}_{1,2}^2 \ll M_V^2$. In that limit, all integrals become trivial and we find~\cite{Liu:2019tuy}
\begin{align}
\chi_{\mathrm{nf}}^H (\vbone,\vbtwo) = 1 - \frac{\pi^2}{3}\,.
\end{align}

In the case of double Higgs production, the form of the corrections is
rather cumbersome but still entirely straightforward and we prefer to
avoid writing down the formulas explicitly. On the other hand, if we
consider the same limit as above, ie $\bm{q}_{1,2}^2 \sim
\bm{q}_{3,4}^2 \ll M_V^2$,  with the additional
   assumption that the Higgs bosons are produced with zero
  rapidity, the formulae simplify considerably. In order to present
the result, we divide the LO cross-section in three contributions as
\begin{align}
d \sigma^{\mathrm{LO}}_{HH} = \mathrm{d}\sigma_{TT}^{\mathrm{LO}} + 
\mathrm{d}\sigma_{BB}^{\mathrm{LO}} + \mathrm{d}\sigma_{TB}^{\mathrm{LO}}, \label{eq:xsHHLO}
\end{align}
where $\mathrm{d}\sigma_{TT}^{\mathrm{LO}}$ is the contributions stemming solely from diagrams $T_1$ and $T_2$,
$\sigma_{BB}^{\mathrm{LO}}$ from $B_1$ and $B_2$ and $\sigma_{TB}^{\mathrm{LO}}$ from the interference of the two classes of diagrams, see Fig.~\ref{fig:feyn-VBFHH}.
 With this, we find that the non-factorisable corrections at NNLO take the suggestive form
 \begin{strip}
\vspace{0.3cm}
\begin{alignat}{3}
  \text{d}\sigma_{HH,\text{nf}}^{\text{NNLO}}  &\sim \tilde{\alpha}_s^2&&\Bigg[
  \left(1-\frac{\pi ^2}{3}\right)\left( \text{d}\sigma_{TT}^{\text{LO}} + \text{d}\sigma_{TB}^{\text{LO}} + \text{d}\sigma_{BB}^{\text{LO}}\right) + \left(z^2-\frac12 z^4 \log^2\left( y  \right)-z^4 \log \left( y  \right) \right) \text{d}\sigma_{TB}^{\text{LO}} 
  \nonumber \\
  & &&+ \Bigg(z^4+2 z^2-2 z^4 \left(1+z^2\right) \log
  \left( y \right)   +z^4\left(z^4-1\right) \log^2\left( y \right)\Bigg)\text{d}\sigma_{BB}^{\text{LO}}\Bigg] \nonumber \\
  & = -\tilde{\alpha}_s^2&&\Bigg[
  2.2899 \cdot \text{d}\sigma_{TT}^{\text{LO}} + 2.2157 \cdot \text{d}\sigma_{TB}^{\text{LO}} +2.1002 \cdot\text{d}\sigma_{BB}^{\text{LO}}\Bigg]\,,
  \label{eq:nnlo-nf_HH_small}
\end{alignat}
\end{strip}
where in this approximation $\mu_V^2 \sim M_V^2 + M_H^2$, and we have defined 
$y=\frac{1+z^2}{z^2}$ and $z=\frac{M_V}{M_H}$.
For the numerical evaluation we have
used $M_V=M_W=80.398~\mathrm{GeV}$ and $M_H=125~\mathrm{GeV}$.
Eq.~\eqref{eq:nnlo-nf_HH_small} shows that the three contributions to
the Born cross-section for di-Higgs production can receive radiative
corrections which are different at the $10\%$ level. The cross-section
for HH production at LO is the result of delicate cancellations of
more than one order of magnitude between the three different
contributions in eq.~\eqref{eq:xsHHLO}, as can be seen in
table~\ref{tab:HH-contributions}.  These cancellations are a well known
manifestation of the role that the Higgs boson has in restoring
unitarity in the Standard Model.  Since we are working in the eikonal
approximation, one could wonder whether this approximation
could spoil these cancellations and induce in this way artificially
large NNLO QCD corrections on the di-Higgs cross-section.  
As a matter of fact, as we will see in detail below, 
the corrections conspire so to preserve the unitarity induced cancellations 
among the various components and produce rather small effects
both at the inclusive and at the differential level.

\begin{figure}
  \centering
  \includegraphics[width=0.46\textwidth,page=1]{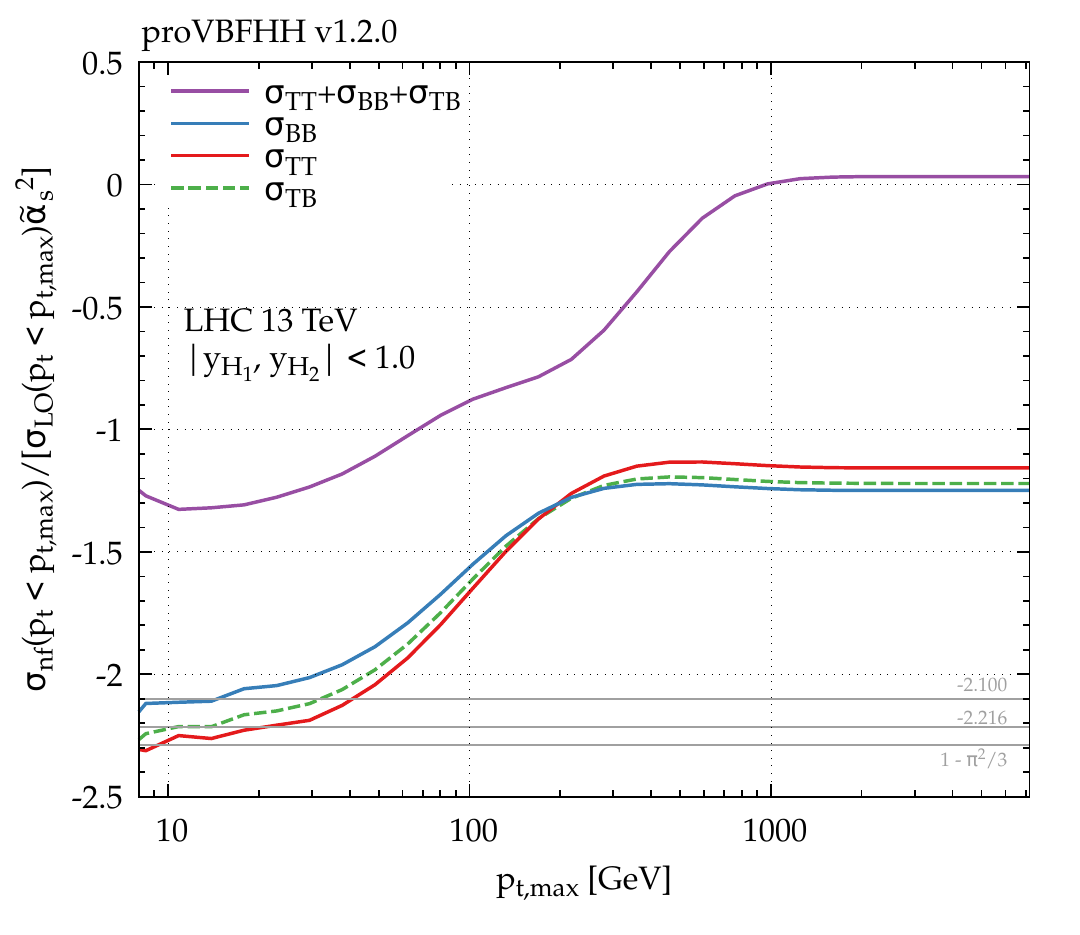}
  \caption{$\tilde{\alpha}_s^2$ coefficient of the $TT$, $TB$ and $BB$ charged-current
    contributions to the NNLO non-factorisable corrections, normalised
    to the corresponding LO term, as a function of a cut on the
    maximum transverse momentum,
    $\max(p_{t,j_1},p_{t,H_1}) < p_{t,\text{max}}$. } 
  \label{fig:expansion-xsct}
\end{figure}

In figure~\ref{fig:expansion-xsct}, we show the $\tilde{\alpha}_s^2$
coefficient of each of the $TT$, $TB$ and $BB$ contributions as a
function of a cut on the maximum transverse momentum,
$\max(p_{t,j_1},p_{t,H_1}) < p_{t,\text{max}}$,
with the only requirement that $|y_{H,1}, y_{H_2}|<1$. For technical 
reasons this plot only includes the charged-current sub-process.
At small values of $p_{t,\text{max}}$ we can observe a convergence of
these coefficients to the analytic expression given in
eq.~(\ref{eq:nnlo-nf_HH_small}).
Above $p_{t,\text{max}}\sim M_W$ there is a transition to different
numerical values of the coefficients, leading to 
a potential spoliation of the cancellation present in the LO cross section.

Another interesting limit to study is the case where the Higgs bosons'
transverse momenta are small, i.e.
$ \bm{q}_3^2,\bm{q}_4^2 \ll M_V^2$ which implies that
$\bm{q}_1^2 \sim \bm{q}_2^2$. In this limit, assuming again that
the Higgs bosons are produced with zero rapidity, it is easy to show that the cross section becomes
\begin{alignat}{3}
  \text{d}\sigma_{HH,\text{nf}}^{\text{NNLO}}  &\sim \tilde{\alpha}_s^2&&\Bigg[\left(1-\frac{\pi ^2}{3}\right)\left( \text{d}\sigma_{TT}^{\text{LO}} + \text{d}\sigma_{TB}^{\text{LO}} + \text{d}\sigma_{BB}^{\text{LO}}\right) \nonumber \\
    & && + C_{TT}  \text{d}\sigma_{TT}^{\text{LO}} + C_{TB}  \text{d}\sigma_{TB}^{\text{LO}}+ C_{BB}  \text{d}\sigma_{BB}^{\text{LO}}\Bigg],
  \label{eq:small-HH-limit}
\end{alignat}
where the coefficients $C_{TT}$, $C_{TB}$ and $C_{BB}$ 
now depend on the variable $x = \bm{q}_1^2/M_V^2 = \bm{q}_2^2/M_V^2$,
in addition to $y$ defined above. As their analytical expression is rather lengthy,
we prefer to provide them  in appendix~\ref{sec:coeffs}. 
Here we notice
that this result reproduces eq.~\eqref{eq:nnlo-nf_HH_small} when
$x \rightarrow 0$, as expected.
On the other hand, as $x$ grows the radiative corrections
to the three contributions differ widely and in particular we see that
as $x\rightarrow \infty$ the cross section takes the form

\begin{equation}
  \text{d}\sigma_{HH,\text{nf}}^{\text{NNLO}}  \sim \tilde{\alpha}_s^2\Bigg[
     x^2 \text{d}\sigma_{TT}^{\text{LO}} + 0.2034 x^3 \text{d}\sigma_{TB}^{\text{LO}} +0.04137 x^4 \text{d}\sigma_{BB}^{\text{LO}}\Bigg].
\end{equation}

In the limit where $\bm{q}_1^2\sim \bm{q}_2^2 \sim M_V^2$, ie when $x=1$ we obtain the numerical expression
\begin{equation}
   \text{d}\sigma_{HH,\text{nf}}^{\text{NNLO}}  \sim \tilde{\alpha}_s^2\Bigg[
  1.1277 \cdot \text{d}\sigma_{TT}^{\text{LO}} + 1.8556 \cdot \text{d}\sigma_{TB}^{\text{LO}} +2.6428 \cdot\text{d}\sigma_{BB}^{\text{LO}}\Bigg].
\end{equation}
This leads us to conclude that the non-factorisable corrections 
can grow very
rapidly with $x$, ie with the transverse momentum of the jets. We note
that the eikonal approximation is strictly speaking only valid when
$x\sim 0$ and that the rapid growth could be seen (at least in part) as a consequence of the breakdown
of the approximation. 

Before concluding this section, it is worth noting that at
$\mathcal{O}(\alpha_s^2)$ there are both loop-induced and real
emission diagrams that contribute to the non-factorisable corrections
discussed above. Nevertheless, it is well known that real emission
diagrams do not contribute to leading order in the eikonal
approximation, and the whole cross-section in this limit stems from
the virtual contributions only.  This is also demonstrated by the fact
that IR divergences cancel between the two-loop and the one-loop
squared amplitudes.  We stress here that the real emission diagrams
have been computed for the single Higgs case
in~\cite{Campanario:2013fsa}, and could be used to compute
non-factorisable corrections beyond the leading eikonal approximation,
once the full two-loop amplitudes become available.

\section{Results for single Higgs VBF production}
\label{sec:results}

\subsection{Setup}
\label{sec:setup}
In order to investigate the size of the various QCD corrections, we
study 13 TeV proton-proton collisions, in a setup identical to Ref.~\cite{Cacciari:2015jma}.
We use a diagonal CKM matrix, full Breit-Wigners for the $W$, $Z$ and
the narrow-width approximation for the Higgs boson.
We take the NNPDF 3.0 parton distribution functions at NNLO with
$\alpha_s(M_Z) = 0.118$
(\texttt{NNPDF30\_nnlo\_as\_0118})~\cite{Ball:2014uwa}, as implemented
in {\texttt{LHAPDF-6.1.6}}~\cite{Buckley:2014ana}.
We consider five light flavours and ignore contributions with top quarks
in the final state or internal lines.
We set the Higgs mass to $M_H = 125\GeV$, compatible with the
experimentally measured value~\cite{Aad:2015zhl}.  Electroweak
parameters are set according to known experimental values and
tree-level electroweak relations. As inputs we use $M_W = 80.398\GeV$,
$M_Z = 91.1876\GeV$ and $G_F = 1.16637\times 10^{-5} \GeV^{-2}$. For
the widths of the vector bosons we use $\Gamma_W = 2.141 \GeV $ and
$\Gamma_Z = 2.4952 \GeV$.
The central factorisation, $\muF$, and renormalisation, $\muR$, scales
are set to
\begin{equation}
  \label{eq:scale}
  \mu_0^2(p_{t,H}) = \frac{M_H}{2} \sqrt{\left(\frac{M_H}{2}\right)^2 +
        p_{t,H}^2}\,,
\end{equation}
when computing factorisable corrections.
We compute the residual scale uncertainties by varying this scale up
and down by a factor 2 keeping $\muR=\muF$, which was shown in
Ref.~\cite{Cacciari:2015jma} to encompass almost the same scale
uncertainty bands as a full 7-point scale variation (i.e.\ where
$\muR$ and $\muF$ are varied independently by a factor 2 with
$\frac{1}{2} \leq \frac{\muR}{\muF} \leq 2$).
For the purpose of comparing these effects, we compute the
non-factorisable corrections using the same central scale, which
differs  from the renormalisation scale choice
$\muR=\sqrt{p_{t,j_1}p_{t,j_2}}$ in Ref.~\cite{Liu:2019tuy}.
The residual scale uncertainties for these last predictions have been
obtained using the full 7-point scale variation.

In the following we will discuss results both fully inclusively in the
VBF jets, and under a set of representative VBF selection cuts. To
pass our VBF selection cuts, events should have at least two jets with
transverse momentum $p_t > 25\GeV$; the two hardest (i.e.\ highest
$p_t$) jets should have absolute rapidity $|y|<4.5$, be separated by a
rapidity $\Delta y_{j_1,j_2} > 4.5$, have a dijet invariant mass
$m_{j_1,j_2} > 600\GeV$ and be in opposite hemispheres
($y_{j_1} y_{j_2} < 0$). 
We define jets using the anti-$k_t$
algorithm~\cite{Cacciari:2008gp}, as implemented in \texttt{FastJet
  v3.1.2}~\cite{Cacciari:2011ma}, with radius parameter $R=0.4$.

We compute all QCD corrections within the {\texttt{proVBFH}}
framework~\cite{Cacciari:2015jma,Dreyer:2016oyx} which is based on
results presented in
Refs.~\cite{Salam:2008qg,Alioli:2010xd,Nason:2009ai,Jager:2014vna,Vogt:2004mw,Moch:2004xu,Vermaseren:2005qc,Moch:2007rq,Buehler:2011ev}.
As of version 1.2.0, the non-factorisable corrections of
Ref.~\cite{Liu:2019tuy} have also been implemented in
{\texttt{proVBFH}}. We evaluate the integrals of eqs.~\eqref{eq:tri1}-\eqref{eq:box2}
using fourth order Runge-Kutta methods.

\subsection{Validity of the eikonal approximation}
\label{sec:nonfact_validity}
\begin{figure*}[ht!]
  \centering
  \begin{minipage}{0.4\linewidth}
    \includegraphics[width=1.0\textwidth,page=1]{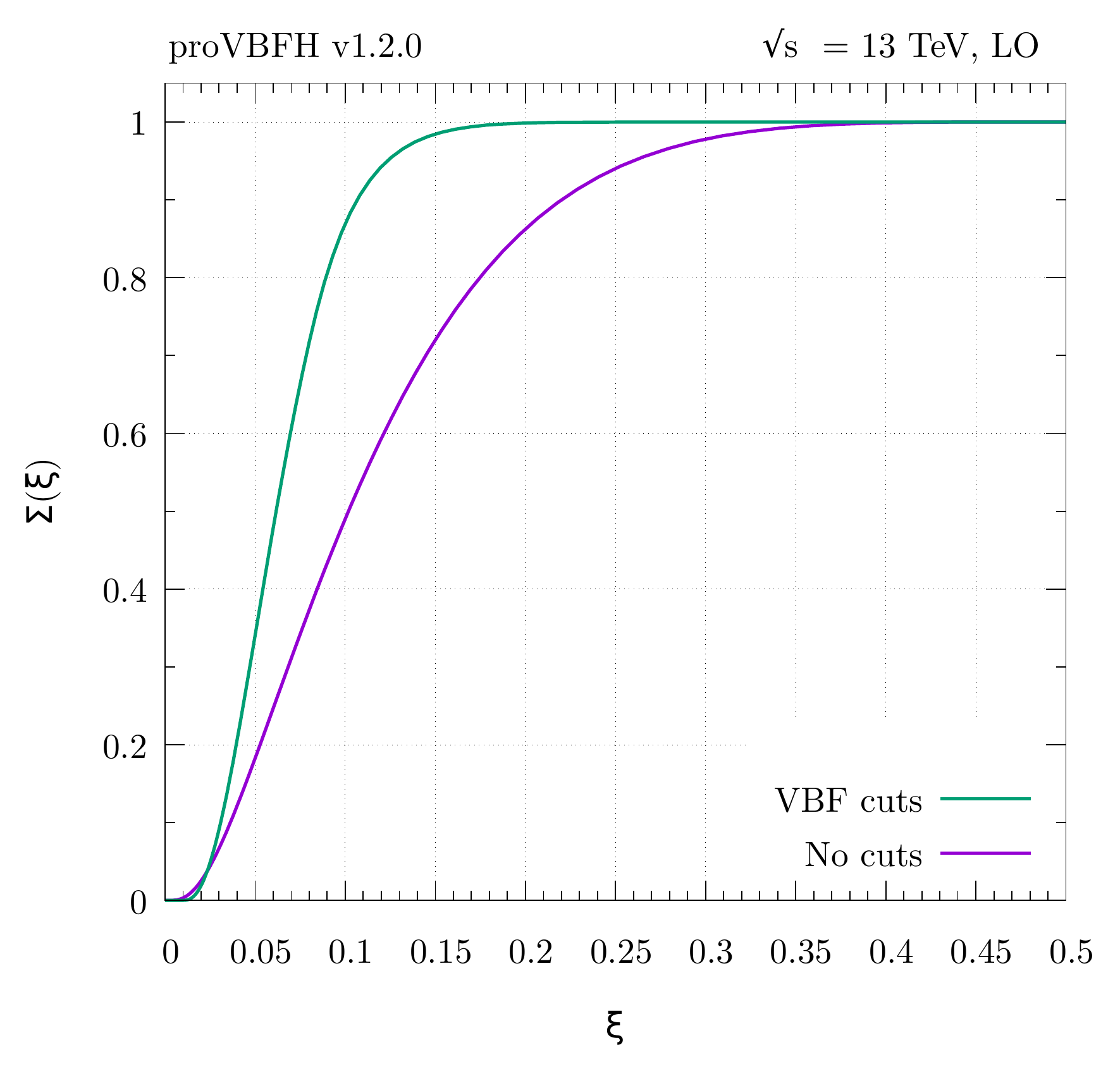}
  \end{minipage}\hspace{0.5cm}
  \begin{minipage}{0.4\linewidth}
  \includegraphics[width=1.0\textwidth,page=1]{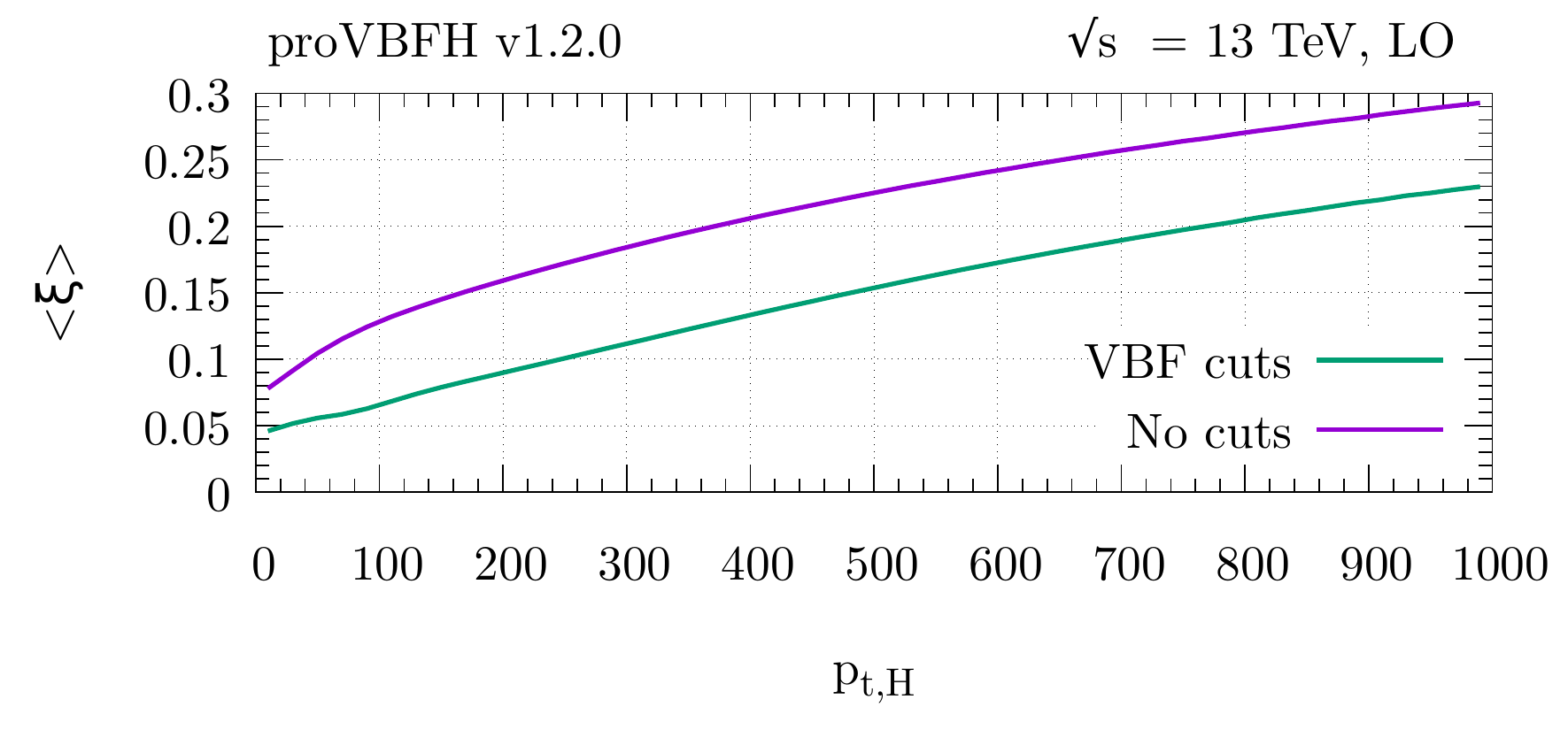}
  \includegraphics[width=1.0\textwidth,page=2]{data/expec-ptjsqrts-paper.pdf}
  \end{minipage}%
  \caption{(left): The normalised integrated cross section as a
    function of $\xi$ fully inclusively (purple) and
    under the VBF cuts of sec. ~\ref{sec:setup} (green) for single
    Higgs production through VBF. (right): The average of
    $\xi$ as a function of $y_H$ and $p_{t,H}$.}
  \label{fig:ptjsqrts-int}
\end{figure*}
The calculation of the non-factorisable NNLO corrections in VBF given
in Ref.~\cite{Liu:2019tuy} is carried out as an expansion in $\xi$
truncated to lowest order in $\xi$, see eq.~\eqref{eq:defxi}. The authors
argue that this ratio is typically of the order $\frac{1}{6}$, based
on experimental measurements of the $p_{t,j_1}$ and $m_{jj}$
spectra~\cite{Khachatryan:2015bnx,Aaboud:2018gay}, and hence that the
relative error associated with truncating the power expansion at the
leading order is roughly $\frac{1}{36}$. This analysis is performed
under the VBF cuts given in sec.~\ref{sec:setup} which guarantee large
$\sqrt{s}$ because of the requirement on the invariant mass of the
di-jet system.

In this section we investigate in some detail how robust this
approximation remains when no cuts are applied to the jets. Although such
an inclusive setup is not of much phenomenological interest, it is of
theoretical interest, given that not only the factorisable NNLO
corrections are known fully inclusively, but also the
N$^3$LO ones~\cite{Dreyer:2016oyx}.

In the left panel of figure~\ref{fig:ptjsqrts-int} we show the
normalised VBF cross section integrated in $\xi$, defined as
\begin{align}
  \Sigma(\xi) = \frac{1}{\sigma}\int_0^{\xi}\frac{d\sigma}{d\xi'}d\xi'\, .
\end{align}
We show the cross section fully inclusively and under VBF cuts. Under
VBF cuts the cross section clearly lives below $\xi\sim 0.2$, whereas
the fully inclusive cross section receives contributions all the way
up to $\xi\sim 0.4$. However almost $85\%$ of the events have
$\xi<0.2$ implying that the approximation used in
Ref.~\cite{Liu:2019tuy} is valid in a large region of the inclusive
VBF phase space. In fact, the average value of $\xi$ is below $0.12$
for all values of the rapidity of the Higgs Boson and for moderate
transverse momenta, $p_{t,H}$, as can be seen in the two right panels
of figure~\ref{fig:ptjsqrts-int}. At large $p_{t,H}$, the average
value of $\xi$ increases almost linearly with $p_{t,H}$ and hence the
eikonal approximation starts to break down. This is not unexpected as
$p_{t,H}$ is balanced by the jet transverse momenta, which by
definition have to take moderate values in order to keep $\xi$ small.

In the region of phase space where the eikonal approximation breaks
down, i.e.\ when $\xi$ becomes large,
\eqref{eq:nnlo-nf_H} is no longer valid.
However, in this region the non-factorisable corrections are not
expected to receive a Glauber phase enhancement partially mitigating the
colour suppression, as this enhancement arises only in the eikonal
limit.

\subsection{Fiducial results}
In this section, we provide results for both factorisable and
non-factorisable corrections on differential and fiducial cross
sections. Although they have been presented separately in
Refs.~\cite{Cacciari:2015jma,Liu:2019tuy}, we show here for the first
time the combined factorisable and non-factorisable NNLO prediction to
VBF Higgs production.

\subsubsection{With VBF cuts}
\label{sec:vbfcuts}
\begin{figure*}
  \centering
  \includegraphics[width=0.33\textwidth,page=1]{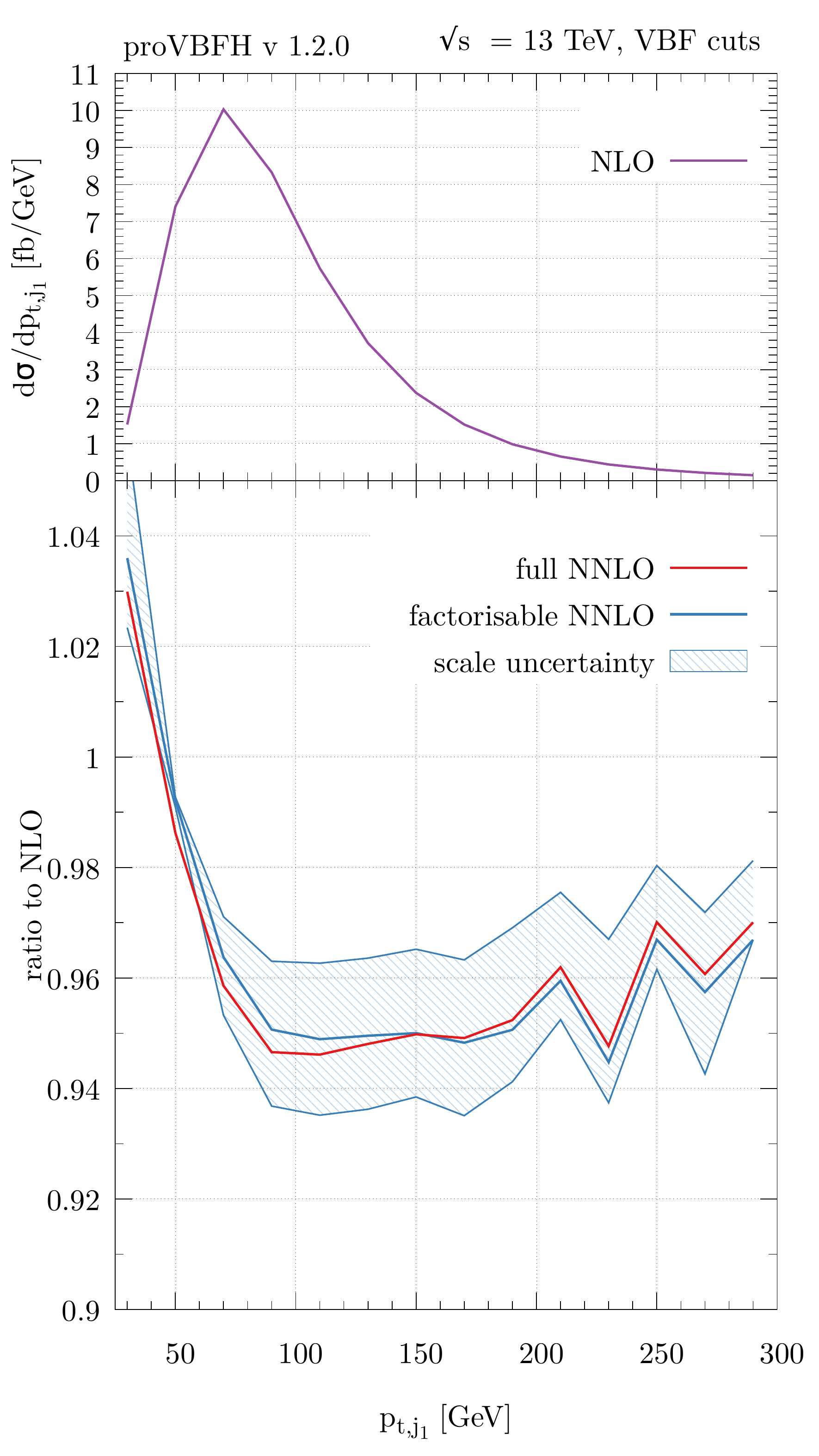}%
  \includegraphics[width=0.33\textwidth,page=2]{data/non-fact-vbf-cuts-paper-new.pdf}%
  \includegraphics[width=0.33\textwidth,page=3]{data/non-fact-vbf-cuts-paper-new.pdf}
  \caption{Upper panel: NLO prediction for VBF production with cuts
    for the transverse momentum of the two leading jets and the Higgs
    boson. Lower panel: Ratio of the factorisable NNLO prediction to
    NLO (blue) and of the full NNLO prediction to NLO (red). The
    blue bands represent the scale uncertainty of the NNLO
    factorisable prediction.}
  \label{fig:pt_vbf}
\end{figure*}
\begin{figure*}
  \centering
  \includegraphics[width=0.33\textwidth,page=7]{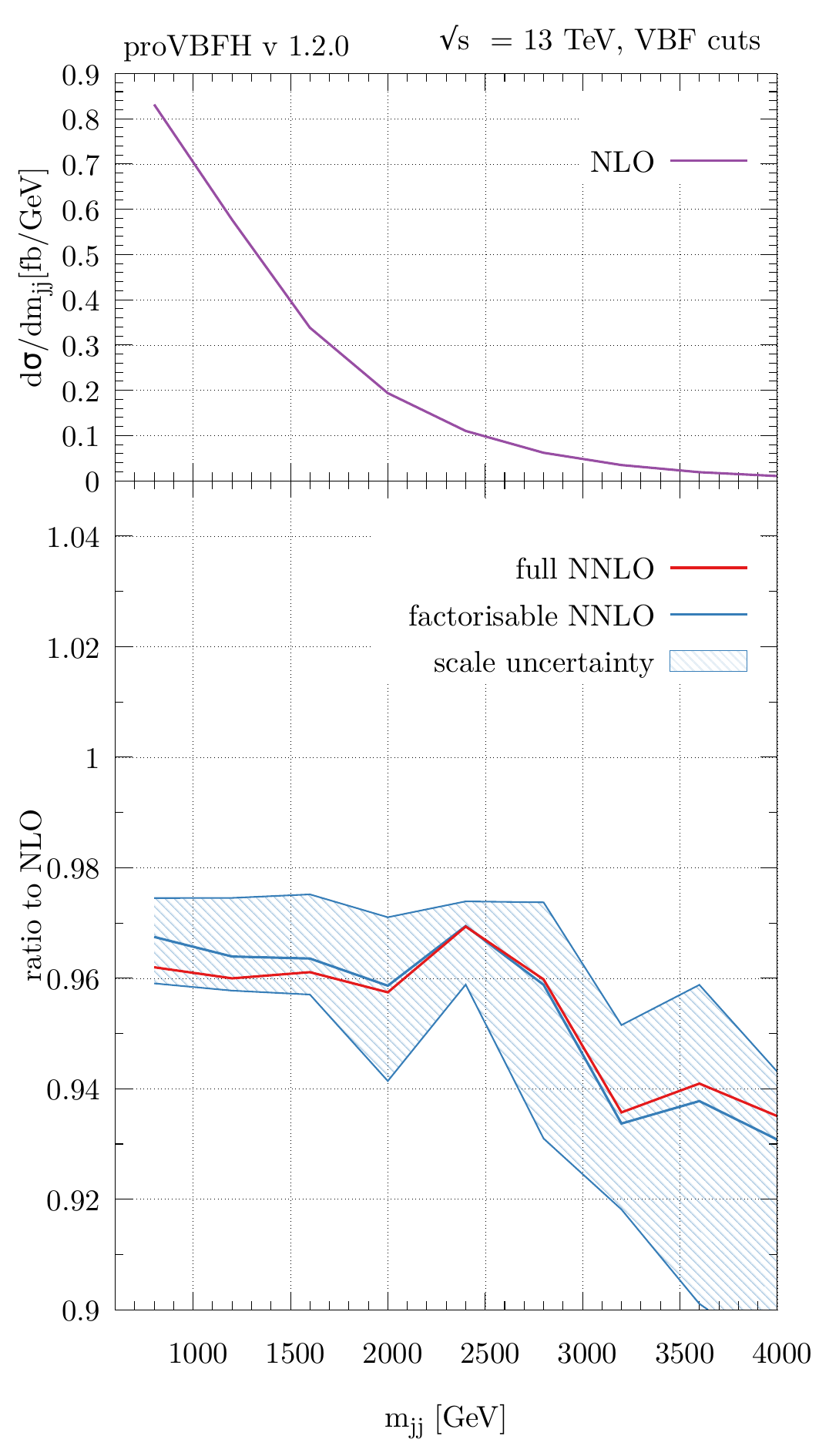}%
  \includegraphics[width=0.33\textwidth,page=8]{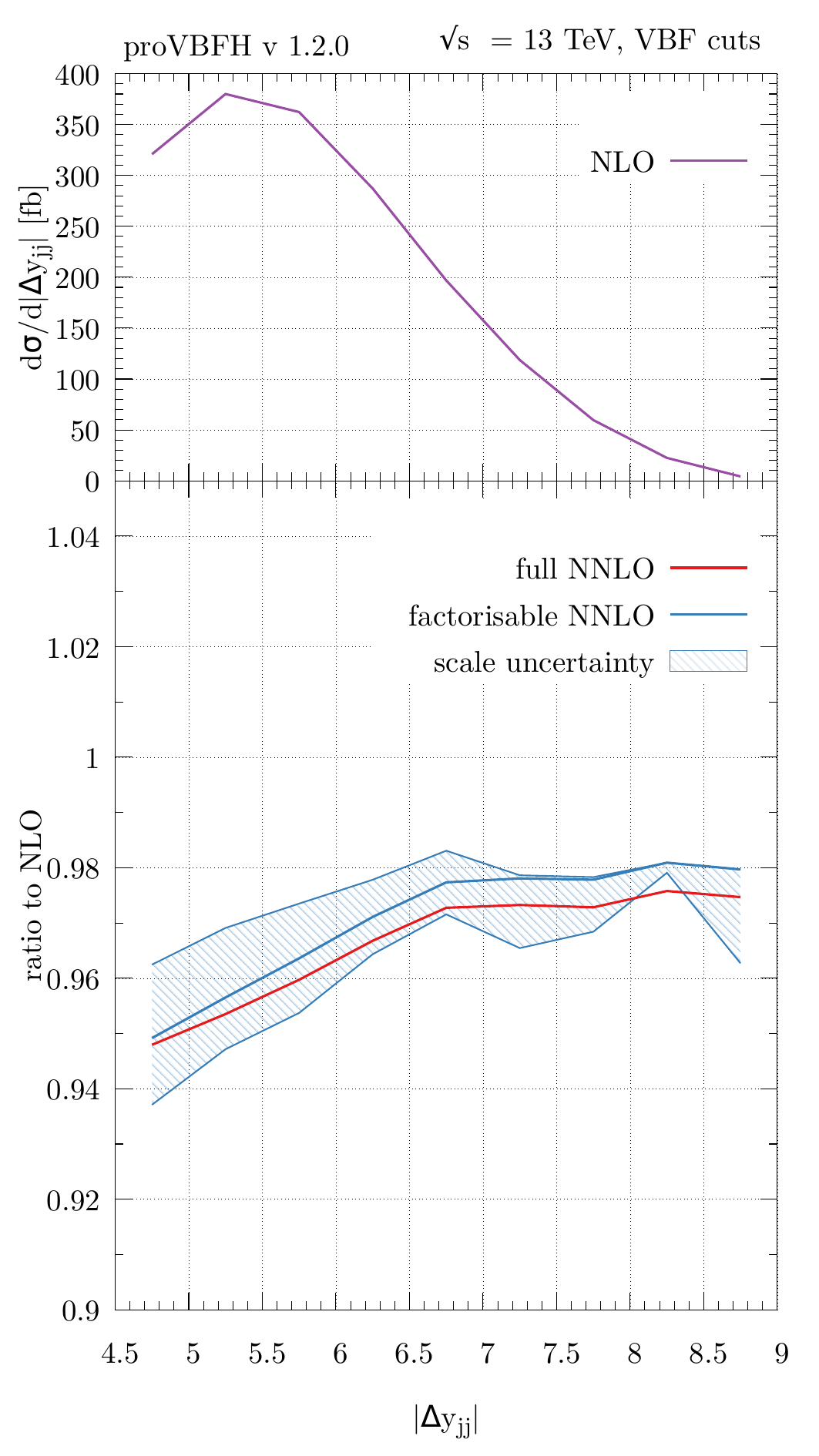}%
  \includegraphics[width=0.33\textwidth,page=6]{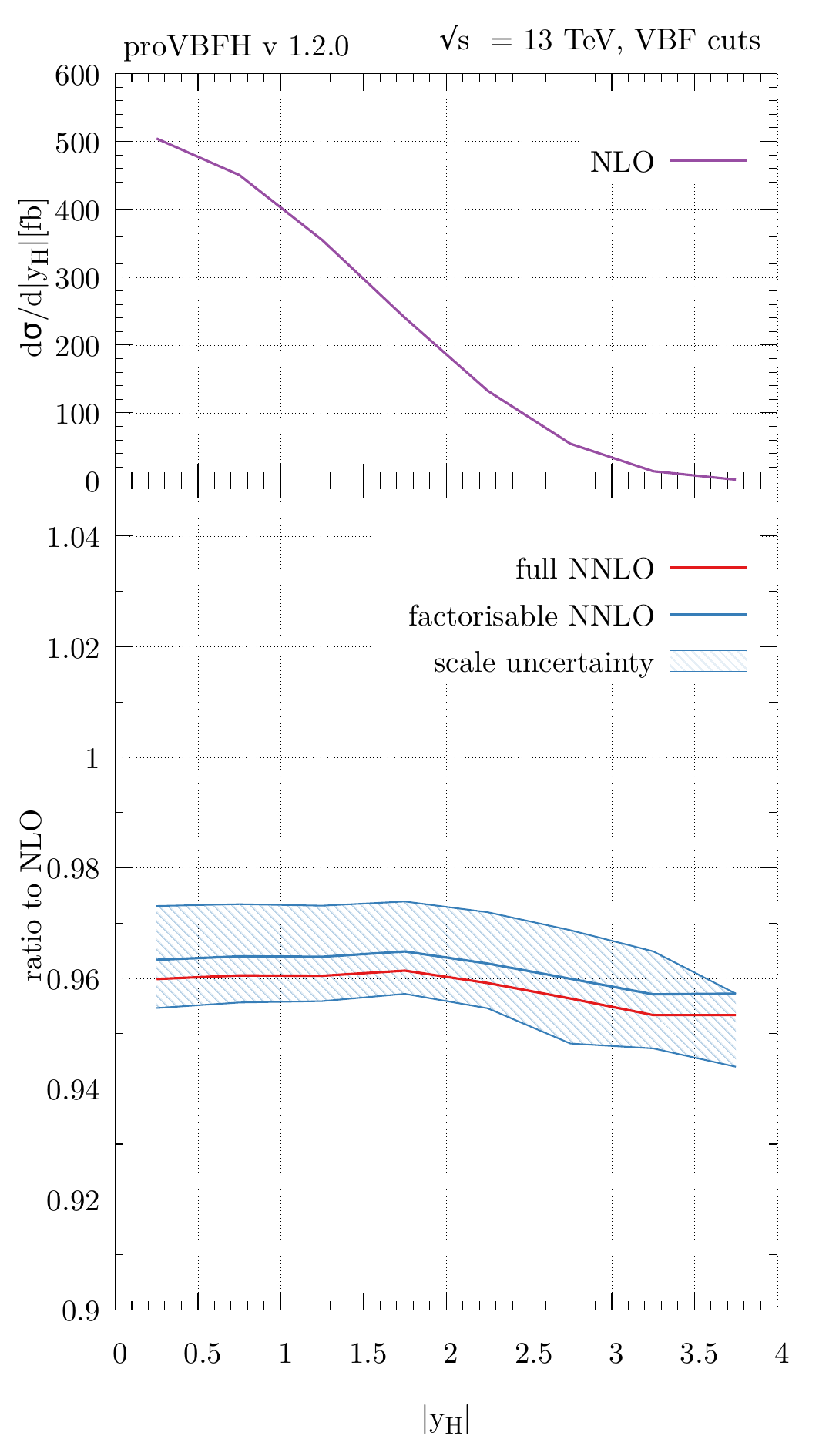}
  \caption{Upper panel: NLO prediction for VBF production with cuts
    for the invariant mass and rapidity separation of the tagging
    jets, as well as for the rapidity of the Higgs boson. Lower panel:
    Ratio of the factorisable NNLO prediction to NLO (blue) and of the
    full NNLO prediction to NLO (red). The blue bands represent the
    scale uncertainty of the NNLO factorisable prediction.}
  \label{fig:jj_vbf}
\end{figure*}
% ----------------------------------------------------------------------
In figures~\ref{fig:pt_vbf} and \ref{fig:jj_vbf} we compare the size
of the factorisable and non-factorisable corrections to VBF Higgs
production under the selection cuts of section~\ref{sec:setup}.
In the upper panels we show the NLO prediction. The lower panels show
various predictions normalised to the NLO prediction.
In blue we show the factorisable NNLO prediction with its associated
scale uncertainty band.
The red curve shows the combined NNLO factorisable and
non-factorisable prediction.
In the bulk of the phase space, the non-factorisable corrections are
small and within the scale uncertainty bands.
However, it is interesting to observe that for large $p_{t,j_2}$ and
$p_{t,H}$ the corrections can in certain regions become larger than
the factorisable scale uncertainty.
This makes the non-factorisable corrections of potential relevance in
boosted Higgs boson searches.
On the other hand it is clear from figure~\ref{fig:ptjsqrts-int} that
the eikonal approximation is not reliable at very high values of
$p_{t,H}$, and the corrections should therefore be applied with care.
A summary of the impact of $\mathcal{O}(\as^2)$ corrections on the
fiducial cross section is shown in figure~\ref{fig:heatmap-fact} as a
function of the $\Delta y_{jj}$ and $m_{jj}$ selection cuts, requiring
also two $R=0.4$ anti-$k_t$ jets with $p_t>25\GeV$ and $|y_j| <
4.5$. The corrections have only a mild dependence on the cuts,
decreasing from roughly $-4\%$ at low cuts to around $-3\%$ at larger
cut values. The non-factorisable corrections shown in
figure~\ref{fig:heatmap-nonfact} on the other hand show a stronger
dependence on the cuts. In general they are suppressed by an order of
magnitude compared to the factorisable corrections. They increase in
size with an increase in the $\Delta y_{jj}$ cut, and decrease as the
$m_{jj}$ cut increases.
The first effect is related to the Glauber enhancement which grows
with the separation of the jets.
The decrease of the non-factorisable corrections as the $m_{jj}$ cut
increases is consistent with figure~\ref{fig:jj_vbf}, where we observe that
these corrections change sign around 2.5 TeV.
It is important to keep in mind that these results will strongly
depend on the choice of jet radius.
Beyond LO the VBF cross section is well-known to be
affected by real radiation escaping the tagging
jets~\cite{Rauch:2017cfu}, while the NNLO non-factorisable corrections
are independent of the jet radius.
Therefore, one should compare the non-factorisable contribution not
only to their factorisable counter-part, but also to the size of the
scale uncertainty bands, particularly for large $R$ values when the
NNLO factorisable corrections become numerically small but their scale
uncertainty remains large.

\subsubsection{Without selection cuts}
\label{sec:nocuts}
\begin{figure*}
  \centering
  \includegraphics[width=0.45\textwidth,page=1]{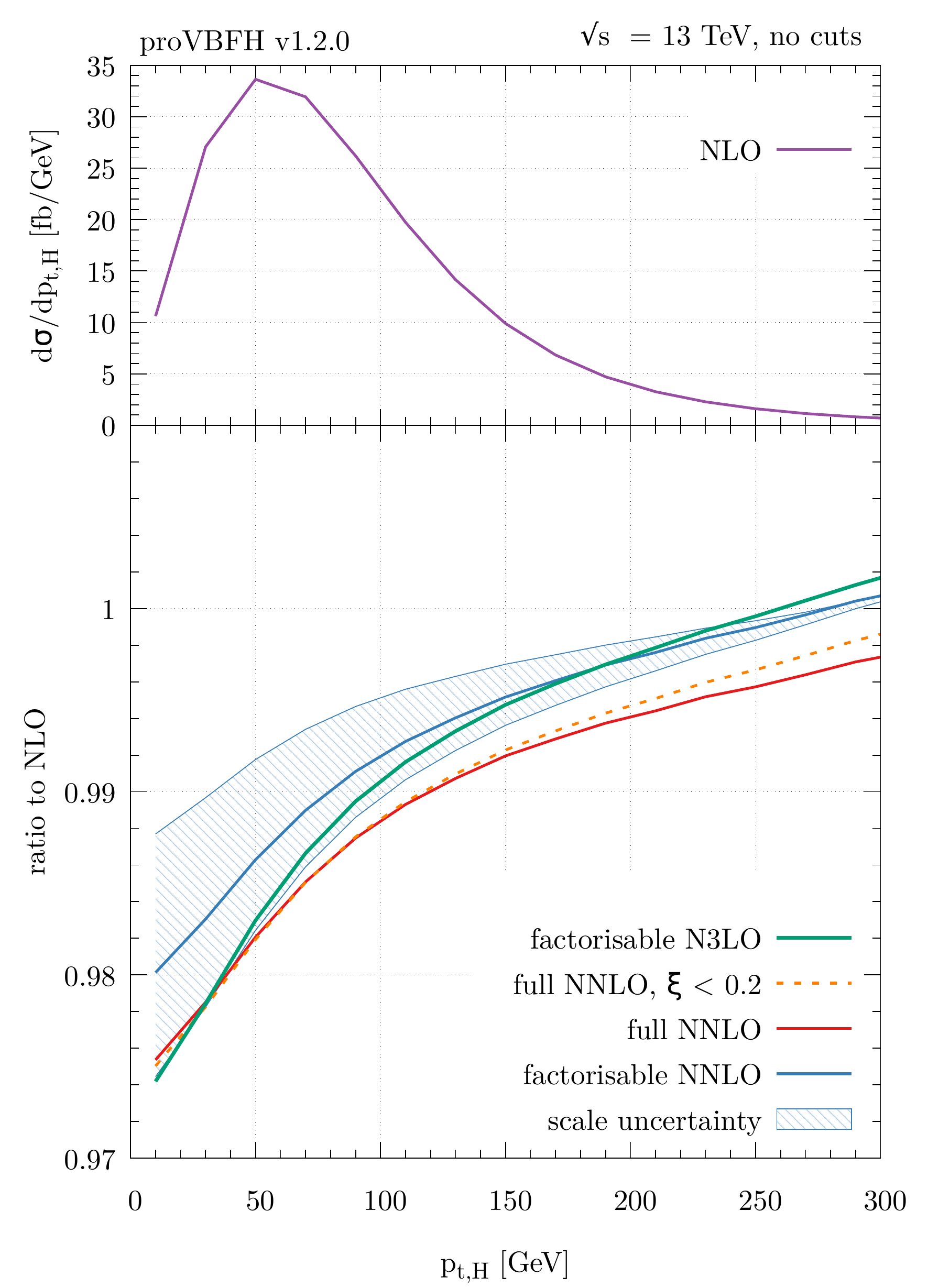}
  \includegraphics[width=0.45\textwidth,page=2]{data/non-fact-incl-paper-new.pdf}
  \caption{Upper panel: NLO prediction for inclusive VBF production
    for the transverse momentum and rapidity of the Higgs boson. Lower
    panel: Ratio of the factorisable NNLO prediction to NLO (blue) and
    of the full NNLO prediction to NLO (red). The blue bands represent
    the scale uncertainty of the NNLO factorisable prediction. The
    orange dashed curve shows the full NNLOprediction after applying a
    cut $\xi<0.2$ on the non-factorisable component. The factorisable
    N$^3$LO prediction is shown in green.}
  \label{fig:n3lo}
\end{figure*}
As discussed in section~\ref{sec:nonfact_validity} the eikonal
approximation is only formally valid when considering
fiducial VBF production.
It can however still provide a useful estimate of the size of the
non-factorisable corrections in inclusive production.
This is of particular interest, as the inclusive factorisable N$^3$LO
corrections are available for comparison in this regime.

In figure~\ref{fig:n3lo} we show in the lower panels for $p_{t,H}$ and
$|y_H|$ in blue the factorisable NNLO prediction, and in green the
factorisable N$^3$LO prediction, both normalised
to the NLO curve shown in the upper panel.
We also show the combined factorisable and non-factorisable NNLO
prediction in two setups: the non-factorisable corrections computed
according to eq.~\eqref{eq:nnlo-nf_H} everywhere in phase
space (red) and the non-factorisable corrections computed according to
eq.~\eqref{eq:nnlo-nf_H} when $\xi<0.2$ and set to $0$
otherwise (dashed-orange).
This last procedure is used to verify that differential observables do
not receive significant contributions from the large $\xi$ region.
We observe that in the bulk of the phase space the numerical
difference between the red and the dashed-orange curves is very
small - of the order of a few percent.
It is therefore clear that the bulk of the non-factorisable
corrections in the eikonal approximation come from the $\xi<0.2$
region, even without VBF cuts.
This is consistent with figure~\ref{fig:ptjsqrts-int} which shows that
the mean value of $\xi$ is typically below $0.2$.
We observe that the non-factorisable NNLO corrections are typically larger than
the factorisable N$^3$LO ones, and that they are not covered by
the NNLO scale variation uncertainties. In fact, the non-factorisable
NNLO corrections are almost $\mathcal{O}(40\%)$ of the factorisable
ones at this order.
However we stress again that the non-factorisable corrections computed
in the eikonal approximation do not necessarily provide reliable
predictions in the full VBF phase space, as subleading terms can
become relevant.
It should also be noted that this large effect stems not from an
enhancement of the non-factorisable effects, but rather from an
order of magnitude decrease in the factorisable corrections when no
cuts are applied.

\section{Results for di-Higgs VBF production}
\label{sec:dihiggs}
We will now investigate the impact of non-factorisable contributions
to the VBF Higgs pair production process.
The electroweak parameters are set identically to the previous
single Higgs study detailed in~\ref{sec:setup}, with a width
$\Gamma_H=4.030 \MeV$ for the internal Higgs propagator, while the
final state Higgs bosons remain in the narrow-width approximation.
For the factorisable corrections the central renormalisation and
factorisation scales are now set to
\begin{equation}
  \label{eq:scale-dihiggs}
  \mu_0^2(p_{t,HH}) = \frac{m_H}{2} \sqrt{\left(\frac{m_H}{2}\right)^2 +
        p_{t,HH}^2}\,,
\end{equation}
and uncertainties from missing higher orders are again estimated by
varying the scales symmetrically up and down by a factor two, as was
discussed in section~\ref{sec:setup}. For the non-factorisable
corrections we pick the same central scales, but when showing the
residual scale uncertainty envelope, we perform the full 7-point scale
variation, i.e.\ varying independently $\muR$ and $\muF$ by a factor 2
but keeping the ratio to the interval
$\frac{1}{2} \leq \frac{\muR}{\muF} \leq 2$.
The NNLO corrections are calculated with {\tt proVBFHH}
v1.2.0~\cite{Dreyer:2018rfu,Dreyer:2018qbw}.

\subsection{Validity of the eikonal approximation}
\begin{figure*}
  \centering
  \begin{minipage}{0.4\linewidth}
    \includegraphics[width=1.0\textwidth,page=2]{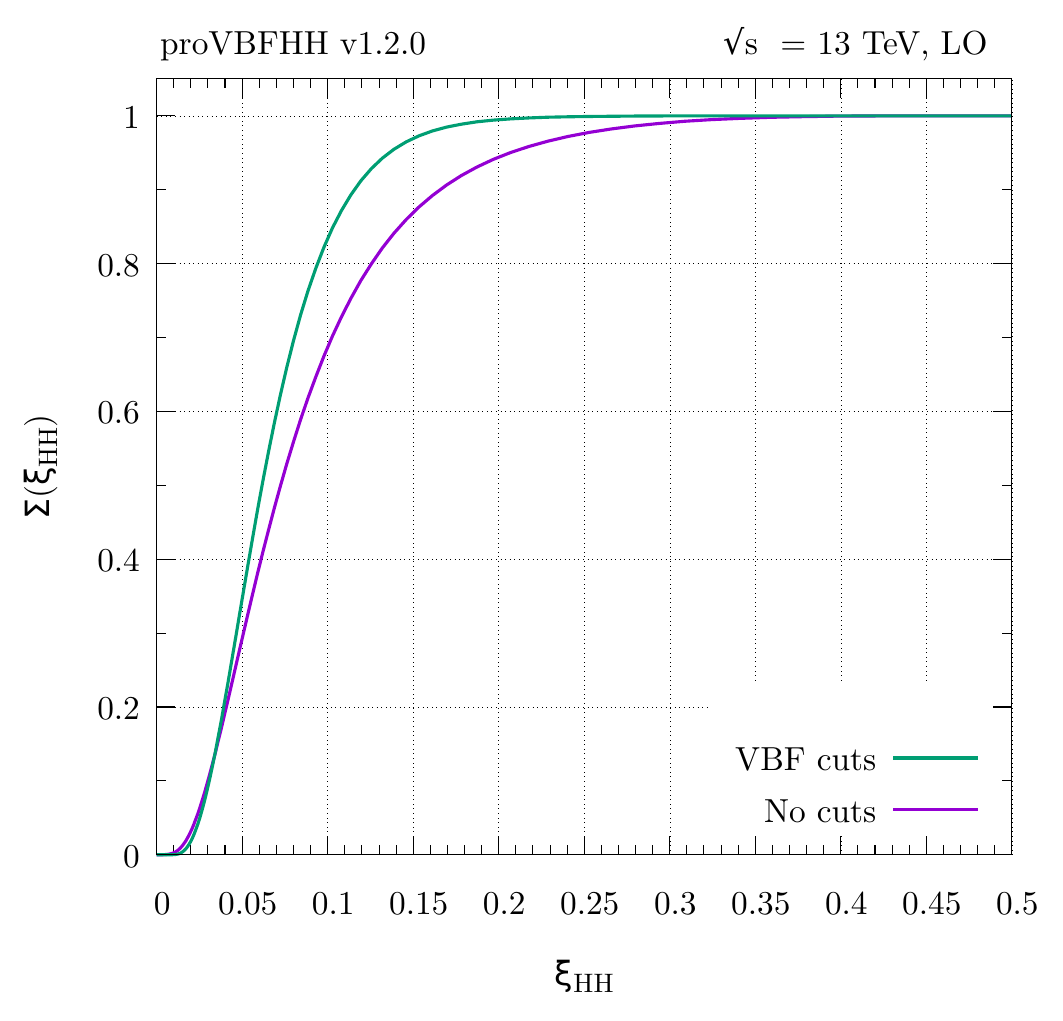}
  \end{minipage}\hspace{0.5cm}
  \begin{minipage}{0.4\linewidth}
  \includegraphics[width=1.0\textwidth,page=3]{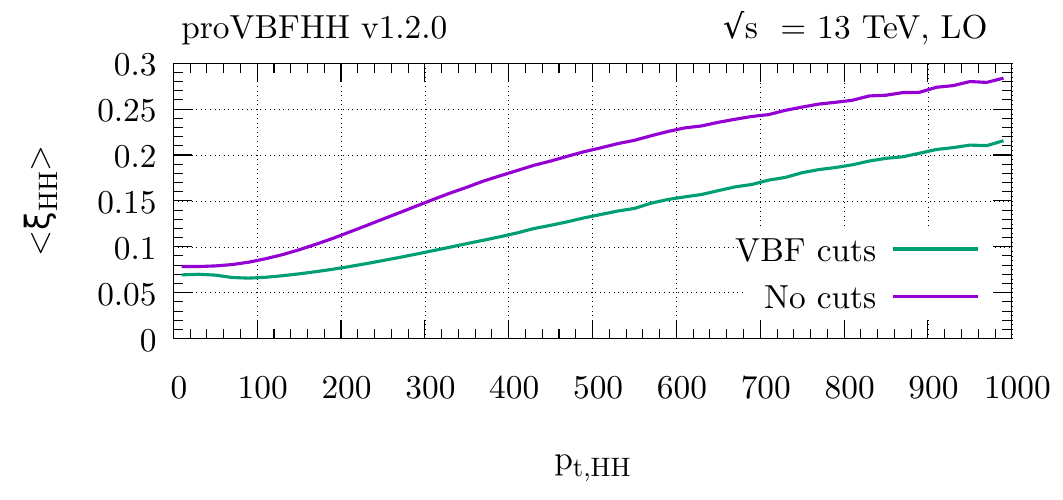}
  \includegraphics[width=1.0\textwidth,page=4]{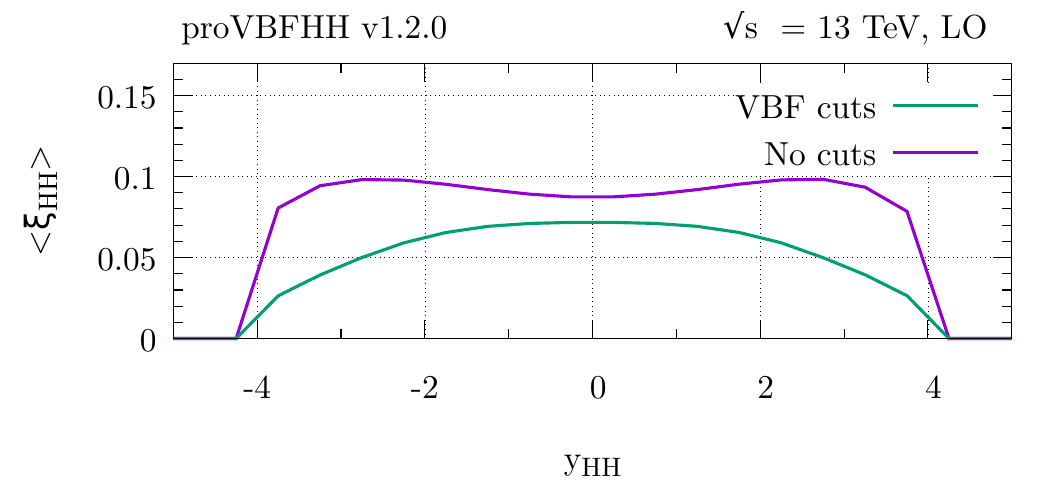}
  \end{minipage}%
  \caption{Left: the normalised integrated cross section as a function
    of $\xi$ fully inclusively (purple) and under the VBF cuts of
    sec. ~\ref{sec:setup} (green) for di-Higgs production through
    VBF. Right: the average of $\xi_{HH}$ as a function of $y_{HH}$ and
    $p_{t,HH}$.}
  \label{fig:ptjsqrts-int-hh}
\end{figure*}

Similarly to what we did for single Higgs production, we start  
by examining the validity of the eikonal approximation.
We expect the eikonal approximation to be valid when all transverse
scales are small compared to the total centre-of-mass energy.
To test this statement quantitatively, we define
\begin{align}
  \xi_{HH} = \frac{\mathrm{max}\{p_{t,j_1},t,u\}}{\sqrt{s}}.
\end{align}
where $t$ and $u$ are defined as below eq.~\eqref{eq:chi-HH}.
In figure~\ref{fig:ptjsqrts-int-hh}, we show in the left panel the
normalised di-Higgs VBF cross section integrated in
$\xi_{HH}$, both fully inclusively and under VBF cuts.
Here we see that compared to the single Higgs process, the $\xi_{HH}$
distribution with no cuts is contained to lower values below
$\xi_{HH}\sim 0.25$. With VBF cuts the two distributions are very
similar, and we therefore expect the eikonal approximation to be valid
also in the di-Higgs process.
In particular, from the right panel of
figure~\ref{fig:ptjsqrts-int-hh}, it is clear that the approximation
only starts to break for very large transverse momentum values of the
Higgs pair. 

\subsection{Fiducial results}
\label{sec:HH-results}
\begin{table}[t] 
  \centering
  \phantom{x}\medskip% add space to make table look nicer...
  \scriptsize{
  \begin{tabular}{lccc|c}
    \toprule
    $\lambda=M_V$ &$\sigma_{TT}$ &$\sigma_{BB}$&$\sigma_{TB}$ & $\Sigma$  \\
    \midrule
    Born          & $10.393$ fb & $14.172$ fb & $-23.904$ fb & $0.662$ fb\\
    1-loop NF     & $0.339\%$   & $0.300\%$   & $0.318\%$    & $0.286\%$  \\
    2-loop NF     & $-0.667\%$  & $-0.621\%$  & $-0.644\%$   & $-0.516\%$ \\
    Full NF       & $-0.327\%$  & $-0.320\%$  & $-0.326\%$   & $-0.230\%$  \\
    \bottomrule
  \end{tabular}}
  \caption{Fiducial cross section for vector-boson fusion di-Higgs
    production at $13\TeV$ under the cuts of sec.~\ref{sec:setup}. The
    first row indicates the Born contribution in femtobarn of the
    triangle diagrams, box diagrams and their interference. The second
    row shows the 1-loop squared non-factorisable (1-loop NF)
    correction in percent of the Born results for the same three
    contributions. The third row shows the same but for the 2-loop
    times tree-level non-factorisable (2-loop NF) contribution. The
    last row shows the same breakdown but for the sum of both
    contributions. The last column shows the sum of the contributions
    across each row.}
\label{tab:HH-contributions}
\end{table}
In this section we discuss the impact of the non-factorisable NNLO
corrections to di-Higgs VBF production computed in
section~\ref{sec:nnlo_nonfact}. As was discussed there, the
non-factorisable corrections are characterised by an interesting interference pattern
which is not present in the single Higgs process. In
table~\ref{tab:HH-contributions} we exemplify this by showing the LO
fiducial cross section under the cuts of section.~\ref{sec:setup} and
their NNLO non-factorisable corrections. We split the cross section
into the contribution coming from only the $T_1$ and $T_2$ topologies,
$\sigma_{TT}$, only the $B_1$ and $B_2$ topologies, $\sigma_{BB}$ and
their interference, $\sigma_{TB}$, c.f.\ figure~\ref{fig:feyn-VBFHH}. As
one can see, the di-Higgs cross section at LO is the result of
cancellations spanning several orders of magnitude.
For the individual sets of diagrams, the non-factorisable corrections are 
below $1\%$ and one can show that the combined 1- and 2-loop contribution in each case is always negative.
It is interesting to note that the relative correction to
$\sigma_{TT}$ of $-0.327\%$ is very close to the correction found in
the single Higgs process of $-0.32\%$ under identical cuts (c.f.\
figure~\ref{fig:heatmap-nonfact}).

To put the size of the NNLO non-factorisable corrections
into context, we compare them to the factorisable NNLO corrections. 
\begin{figure*}[htb]
  \centering
  \subfloat[]{\includegraphics[width=0.5\textwidth,page=1]{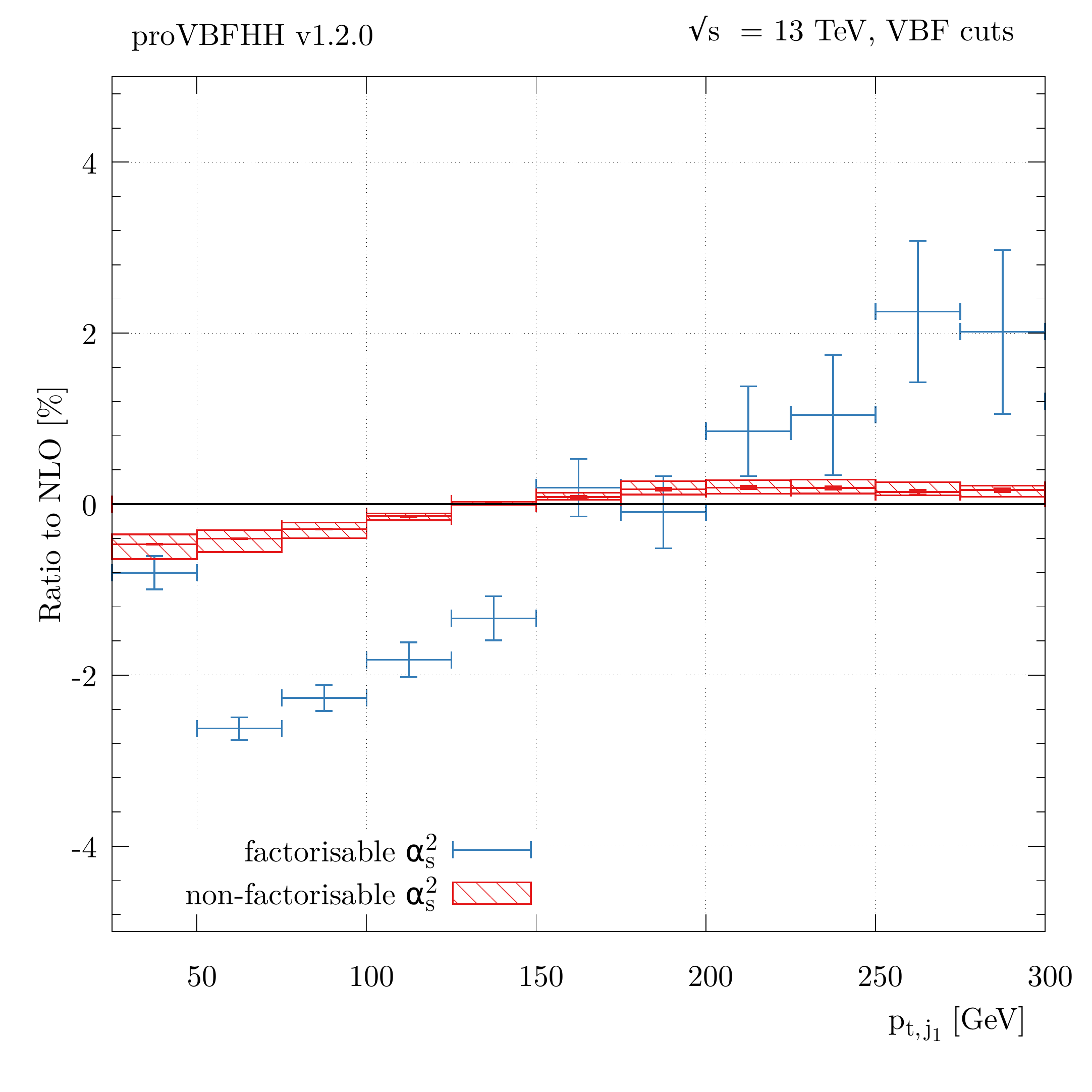}%
  \label{fig:HH-ptj1}}%
  \subfloat[]{\includegraphics[width=0.5\textwidth,page=2]{data/VBFHH-plots-with-scale.pdf}%
  \label{fig:HH-ptj2}}
\caption{Kinematic distributions for Higgs pair production through VBF
  under the cuts of sec.~\ref{sec:setup}. (a) transverse momentum of
  the hardest jet (b) transverse momentum of the second hardest
  jet. In red we show the non-factorisable $\as^2$ correction and in
  blue we show the factorisable one. Both are normalised to the NLO
  cross section.}
  \label{fig:HH-ptj}
\end{figure*}
\begin{figure*}[htb]
  \centering
  \subfloat[]{\includegraphics[width=0.5\textwidth,page=3]{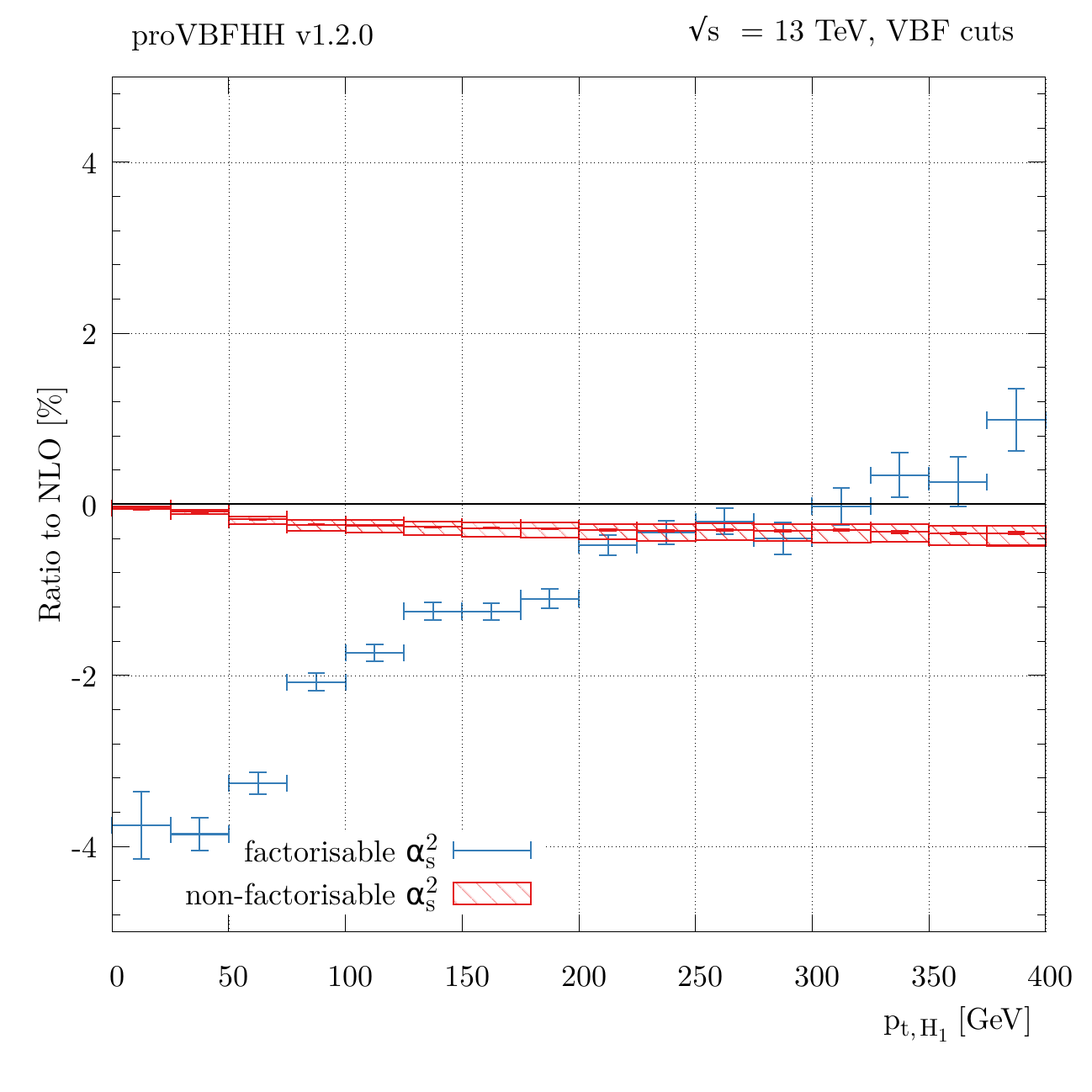}%
  \label{fig:HH-ptH-hard}}%
  \subfloat[]{\includegraphics[width=0.5\textwidth,page=11]{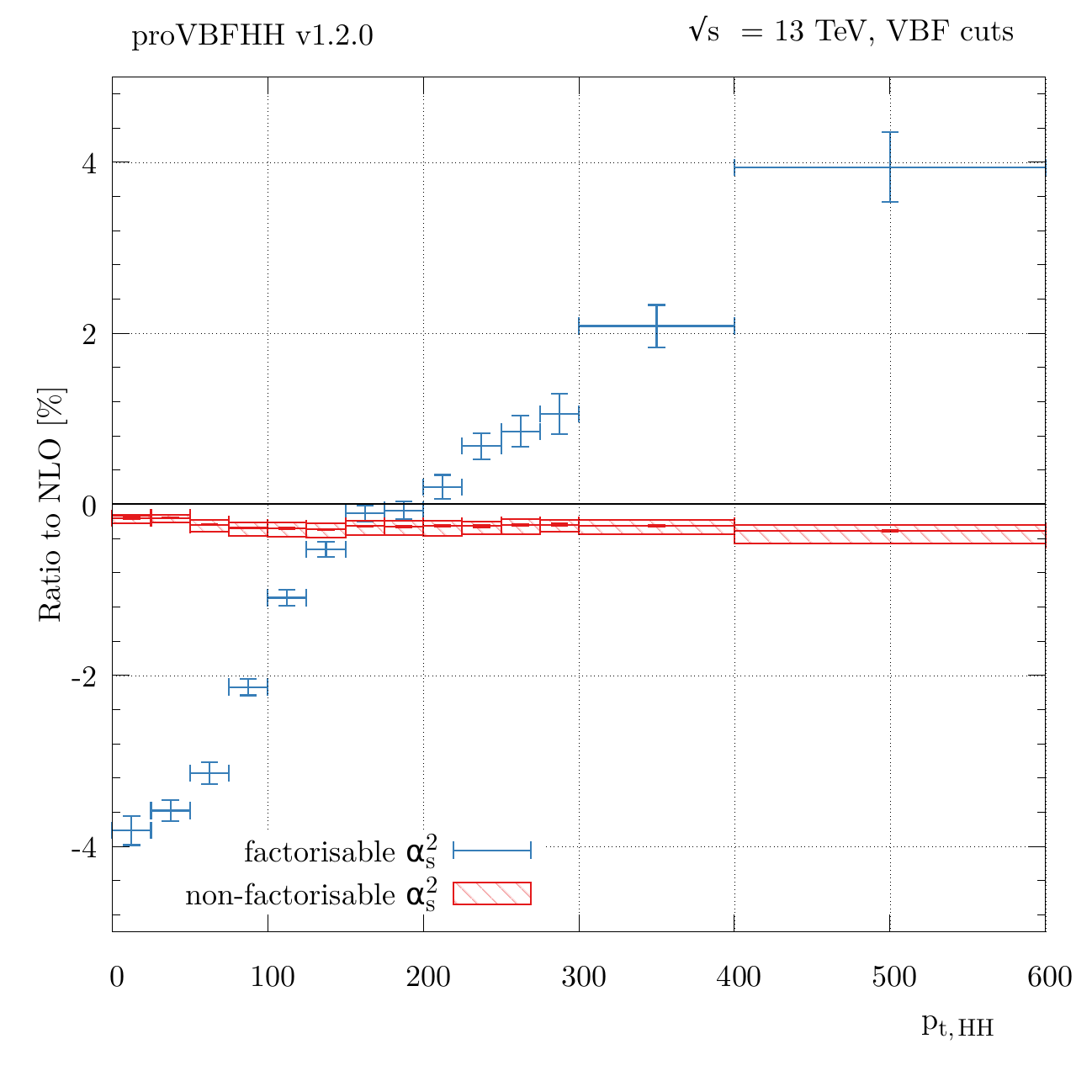}%
  \label{fig:HH-ptHH}}
\caption{Same as figure~\ref{fig:HH-ptj} for (a) transverse momentum of
  the hardest Higgs boson (b) transverse momentum of the di-Higgs
  system.}
  \label{fig:HH-ptH}
\end{figure*}\begin{figure*}[htb]
  \centering
  \subfloat[]{\includegraphics[width=0.5\textwidth,page=12]{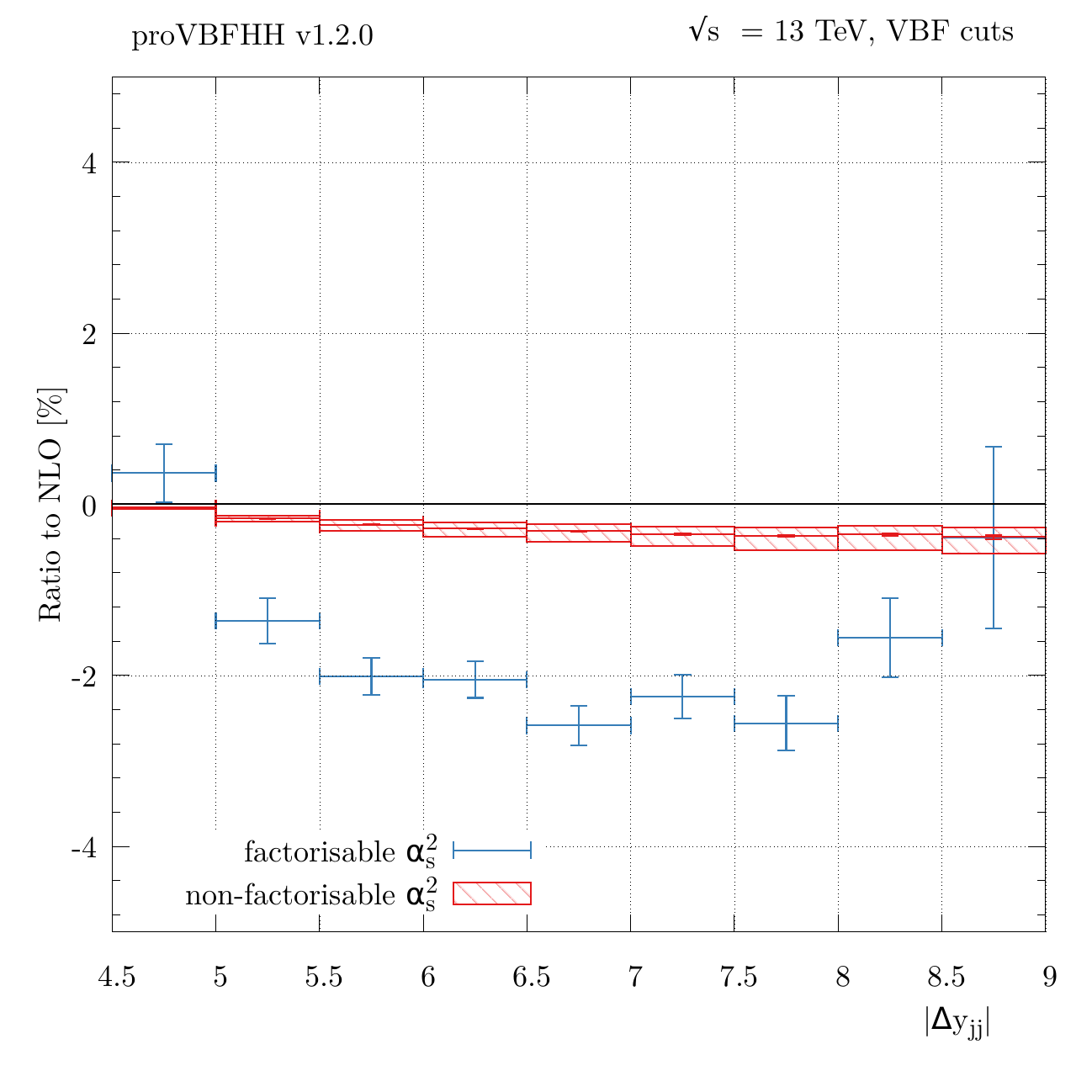}%
  \label{fig:HH-yjj}}%
  \subfloat[]{\includegraphics[width=0.5\textwidth,page=13]{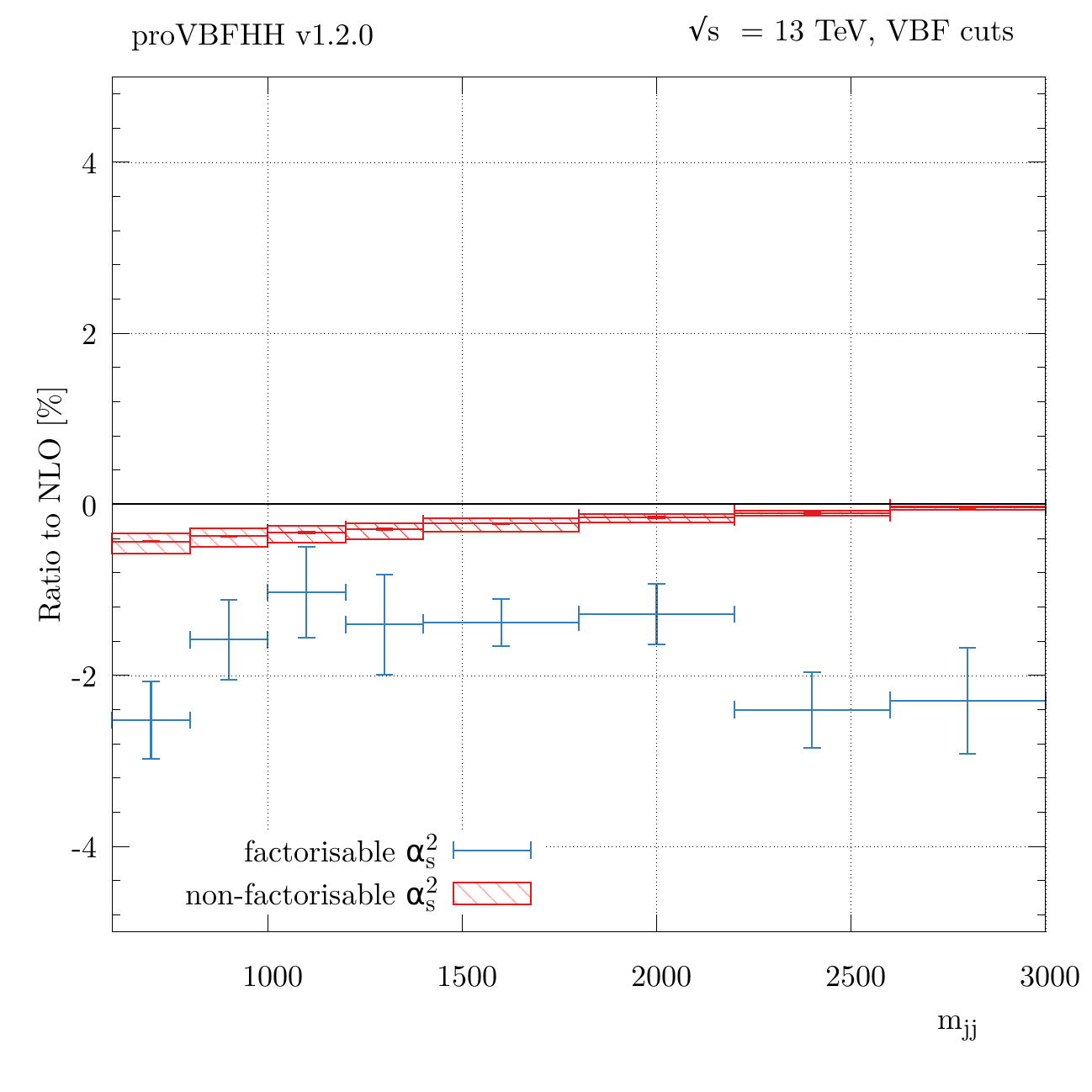}%
  \label{fig:HH-mjj}}
\caption{Same as figure~\ref{fig:HH-ptj} for (a) dijet rapidity
  separation (b) dijet invariant mass.}
  \label{fig:HH-vbf-obs}
\end{figure*}
In figure~\ref{fig:HH-ptj} we show the corrections to the two hardest
jets, normalised to the NLO cross section. For low to moderate jet
transverse momenta the non-factorisable corrections are at the few
permille level and typically smaller in size than their factorisable
counterparts. For $p_{t,j_2}$ they can grow to $1-2\%$ when the transverse momentum becomes
large, see figure~\ref{fig:HH-ptj}. 
We also note here that a similar growth of the NNLO non-factorisable corrections 
in the jet $p_t$ distributions can be observed also in single Higgs production, see
figure~\ref{fig:pt_vbf}.
We should also stress here that, as the jet transverse
momenta grow, the eikonal approximation becomes less reliable and
hence one should use the results in this region with caution.

In figure~\ref{fig:HH-ptH} we show the transverse momentum
distribution of the Higgs with larger transverse momentum and of the
di-Higgs system.
The corrections remain moderate, and tend to be
substantially smaller than the factorisable corrections.

Finally, in figure~\ref{fig:HH-vbf-obs} we show the dijet rapidity
separation and invariant mass. The non-factorisable corrections tend
again at the few permille level in both observables, and smaller than
the factorisable corrections.

\section{Conclusions}
\label{sec:conclusion}
\begin{table}[t] % 
  \centering
  \phantom{x}\medskip% 
  \begin{tabular}{lccc}
    \toprule
    VBF & $\sigma^\text{(NLO fact.)}$  & $\delta\sigma^\text{(fact.)}$ & $\delta\sigma^\text{(non-fact.)}$ \\
    \midrule
    $Hjj$ &  $0.876$    &  $-0.032\,^{+0.008}_{-0.008}$ &  $-0.0030\,^{+0.0006}_{-0.0009}$ [pb] \\
    $HHjj$ &  $0.607$   &  $-0.012\,^{+0.003}_{-0.001}$&  $-0.0015\,^{+0.0004}_{-0.0005}$ [fb]\\
    \bottomrule
  \end{tabular}
  \caption{ Fiducial cross sections for single and double Higgs VBF production at NLO,
    along with the corresponding NNLO corrections.
    The quoted uncertainties correspond to scale dependence, while
    statistical errors at NNLO are about $0.5\permil$. For details on
    the scale variation see sec~\ref{sec:setup}. }
\label{tab:cross-sections}
\end{table}

%----------------------------------------------------------------------
In this paper we have studied the relative sizes of factorisable and
non-factorisable QCD corrections to both single and double VBF Higgs
production.
A summary of the results is given in table~\ref{tab:cross-sections},
which shows the NLO fiducial cross section of single and di-Higgs
production and the corresponding NNLO corrections.
This study was made possible by recent advances in estimating the
non-factorisable terms contributing to the NNLO cross
section~\cite{Liu:2019tuy}, which we extended to the di-Higgs process.
We have presented the combined factorisable and non-factorisable NNLO
corrections, as implemented in the public code
\href{https://provbfh.hepforge.org/}{\texttt{proVBFH} v1.2.0} for
single Higgs VBF production.
We find that for typical selection cuts the non-factorisable NNLO
corrections are small and mostly contained within the factorisable
scale uncertainty bands.
For large jet and Higgs transverse momenta, the non-factorisable
corrections can become comparable to the factorisable ones. In this
region, it is however not clear that the eikonal approximation used
to estimate the corrections remains valid.

We also showed that the corrections computed in
Ref.~\cite{Liu:2019tuy} can be used to provide an estimate for the
non-factorisable corrections for the fully inclusive VBF phase
space.
In this case we find that the non-factorisable NNLO corrections are of
the same order as the NNLO factorisable corrections, and moderately
larger than the factorisable N$^3$LO corrections.
This is in contrast with the usual statement that non-factorisable
corrections can be neglected at this order~\cite{Bolzoni:2010xr} for
inclusive quantities.
We stress that this estimate comes from an extrapolation of the
eikonal approximation into a regime beyond where it is expected to
remain valid, and should therefore only be taken as an estimation of
the true size of non-factorisable NNLO corrections to fully inclusive
VBF.

Finally, we have implemented the non-factorisable correction to the
Higgs pair production process in VBF, which is available in
\href{https://provbfh.hepforge.org/}{\texttt{proVBFHH} v1.2.0}.
We find that the non-factorisable corrections to double-Higgs
production follow similar patterns as were found in the single-Higgs
case. There is however a suppression of the non-factorisable corrections
which comes from a delicate cancellation of the various Born diagrams.

Public versions of the codes used in this article are available
online~\cite{provbfh}.
These results pave the way for precision measurements of the Higgs
sector at the LHC and HL-LHC, as well as for further studies of
non-factorisable effects and their interplay with the choice of jet
radius.

\begin{acknowledgement}
  \textbf{Acknowledgments:} We would like to thank Kirill Melnikov,
  Fabrizio Caola, and Thomas Gehrmann for many enlightening
  discussions.
  We are thankful to Alexander Penin for pointing us to an inconsistency
  in the first version of this paper.
  L.T. wishes to thank Ettore Remiddi and David Kosower for clarifying
  discussions on the reduction of loop integrals in two dimensions.
  We are also grateful to Kirill Melnikov, Gavin Salam, and Giulia
  Zanderighi for useful comments on the manuscript.
  F.D.\ is supported by the Science and Technology Facilities Council
  (STFC) under grant ST/P000770/1.
  A.K.\ is supported by the European Research Council (ERC) under
  the European Union’s Horizon 2020 research and innovation programme
  (grant agreement No. 788223, PanScales), and by Linacre College,
  Oxford.
  A.K. acknowledges support from the Swiss National Science Foundation
  (SNF) under grant number 200020-175595 while part of this work was
  carried out. L.T. is supported by the Royal Society through grant
  URF/R1/191125.
\end{acknowledgement}

\appendix

\newpage
\section{Coefficients}
\label{sec:coeffs}
\begin{strip}
In this appendix we report the coeffcients $C_{TT}$, $C_{TB}$, and
$C_{BB}$ that enter in eq.~\eqref{eq:small-HH-limit}. Defining $x = \bm{q}_1^2/M_V^2 = \bm{q}_2^2/M_V^2$ and $x_\pm = x\pm 1$, they are given by
\begin{align}
  C_{TT} & = -2 \text{Li}_2(-x)+(x-2) x+2 x_+ \log \left(x_+\right)\,, \\
  C_{TB} & = -x_+^2 z^4 \text{Li}_2\left(-\frac{x}{y}\right)+\text{Li}_2(-x) \left(x_+^2 z^4-2\right)-\frac{1}{2} x_+^2 z^4
  \log ^2(y)-2 x_+^2 z^4 \log ^2(x+y) \nonumber\\
  & +2 x_+^2 z^4 \log (y) \log (x+y)-2 x_+^2 z^4 \log \left(x_+\right) \log
  (y)+4 x_+^2 z^4 \log \left(x_+\right) \log (x+y) \nonumber \\
  & -x_- x_+^2 z^4 \log (y)+2 x_- x_+^2 z^4 \log (x+y)-2 x_+^2
  z^4 \log ^2\left(x_+\right)+x_-^2 (x_+) z^2 \nonumber \\
  &+ x_+ \left(x_+ \left(z^2-2 x_- z^4\right)+2\right) \log
  \left(x_+\right)+(x-2) x \,,\\
  C_{BB} & = -2 x_+^2 z^4 \text{Li}_2\left(-\frac{x}{y}\right)+\text{Li}_2(-x) \left(2 x_+^2 z^4-2\right)+x_+^2 z^4
  \left(x_+^2 z^4-1\right) \log ^2(y)\nonumber \\
  & +4 x_+^2 z^4 \left(x_+^2 z^4-1\right) \log ^2(x+y)+4 x_+^2 z^4
  \left(x_+^2 z^4-1\right) \log \left(x_+\right) \log (y) \nonumber \\
  & +\left(8 x_+^2 z^4-8 x_+^4 z^8\right) \log
  \left(x_+\right) \log (x+y)+\left(4 x_+^2 z^4-4 x_+^4 z^8\right) \log (y) \log (x+y) \nonumber \\
  & -2 x_- x_+^2 z^4
  \left(x_+ z^2+1\right) \log (y)+4 x_- x_+^2 z^4 \left(x_+ z^2+1\right) \log (x+y)\nonumber \\
  & +4 x_+^2 z^4 \left(x_+^2
  z^4-1\right) \log ^2\left(x_+\right)+x_-^2 \left(x_+ z^2+1\right){}^2\nonumber \\
  & -2 x_+ \left(x_+ z^2+1\right) \left(2
   x_- x_+ z^4-1\right) \log \left(x_+\right)-1\,.
\end{align}
\end{strip}

% =================================================
\bibliography{nonfact}

\providecommand{\href}[2]{#2}\begingroup\raggedright\begin{thebibliography}{10}

\bibitem{Aad:2012tfa}
{\bf ATLAS Collaboration} Collaboration, G.~Aad {\em et al.}, {\em {Observation
  of a new particle in the search for the Standard Model Higgs boson with the
  ATLAS detector at the LHC}}.
  \href{http://dx.doi.org/10.1016/j.physletb.2012.08.020}{Phys.Lett. {\bf B716}
  (2012)  1--29},
\href{http://arxiv.org/abs/1207.7214}{{\tt arXiv:1207.7214 [hep-ex]}}.
%%CITATION = ARXIV:1207.7214;%%.

\bibitem{Chatrchyan:2012xdj}
{\bf CMS} Collaboration, S.~Chatrchyan {\em et al.}, {\em {Observation of a New
  Boson at a Mass of 125 GeV with the CMS Experiment at the LHC}}.
  \href{http://dx.doi.org/10.1016/j.physletb.2012.08.021}{Phys. Lett. {\bf
  B716} (2012)  30--61},
\href{http://arxiv.org/abs/1207.7235}{{\tt arXiv:1207.7235 [hep-ex]}}.
%%CITATION = ARXIV:1207.7235;%%.

\bibitem{DiMicco:2019ngk}
J.~Alison {\em et al.}, ``{Higgs Boson Pair Production at Colliders: Status and
  Perspectives},'' in {\em {Double Higgs Production at Colliders}},
  B.~Di~Micco, M.~Gouzevitch, J.~Mazzitelli, and C.~Vernieri, eds.
\newblock 9, 2019.
\newblock \href{http://arxiv.org/abs/1910.00012}{{\tt arXiv:1910.00012
  [hep-ph]}}.

\bibitem{Figy:2003nv}
T.~Figy, C.~Oleari, and D.~Zeppenfeld, {\em {Next-to-leading order jet
  distributions for Higgs boson production via weak boson fusion}}.
  \href{http://dx.doi.org/10.1103/PhysRevD.68.073005}{Phys. Rev. {\bf D68}
  (2003)  073005},
\href{http://arxiv.org/abs/hep-ph/0306109}{{\tt arXiv:hep-ph/0306109
  [hep-ph]}}.
%%CITATION = HEP-PH/0306109;%%.

\bibitem{Bolzoni:2010xr}
P.~Bolzoni, F.~Maltoni, S.-O. Moch, and M.~Zaro, {\em {Higgs production via
  vector-boson fusion at NNLO in QCD}}.
  \href{http://dx.doi.org/10.1103/PhysRevLett.105.011801}{Phys. Rev. Lett. {\bf
  105} (2010)  011801},
\href{http://arxiv.org/abs/1003.4451}{{\tt arXiv:1003.4451 [hep-ph]}}.
%%CITATION = ARXIV:1003.4451;%%.

\bibitem{Jager:2014vna}
B.~Jäger, F.~Schissler, and D.~Zeppenfeld, {\em {Parton-shower effects on
  Higgs boson production via vector-boson fusion in association with three
  jets}}. \href{http://dx.doi.org/10.1007/JHEP07(2014)125}{JHEP {\bf 07} (2014)
   125},
\href{http://arxiv.org/abs/1405.6950}{{\tt arXiv:1405.6950 [hep-ph]}}.
%%CITATION = ARXIV:1405.6950;%%.

\bibitem{Cacciari:2015jma}
M.~Cacciari, F.~A. Dreyer, A.~Karlberg, G.~P. Salam, and G.~Zanderighi, {\em
  {Fully Differential Vector-Boson-Fusion Higgs Production at
  Next-to-Next-to-Leading Order}}.
  \href{http://dx.doi.org/10.1103/PhysRevLett.115.082002}{Phys. Rev. Lett. {\bf
  115} (2015) no.~8, 082002},
\href{http://arxiv.org/abs/1506.02660}{{\tt arXiv:1506.02660 [hep-ph]}}.
%%CITATION = ARXIV:1506.02660;%%.

\bibitem{Cruz-Martinez:2018rod}
J.~Cruz-Martinez, T.~Gehrmann, E.~W.~N. Glover, and A.~Huss, {\em {Second-order
  QCD effects in Higgs boson production through vector boson fusion}}.
  \href{http://dx.doi.org/10.1016/j.physletb.2018.04.046}{Phys. Lett. {\bf
  B781} (2018)  672--677},
\href{http://arxiv.org/abs/1802.02445}{{\tt arXiv:1802.02445 [hep-ph]}}.
%%CITATION = ARXIV:1802.02445;%%.

\bibitem{Dreyer:2016oyx}
F.~A. Dreyer and A.~Karlberg, {\em {Vector-Boson Fusion Higgs Production at
  Three Loops in QCD}}.
  \href{http://dx.doi.org/10.1103/PhysRevLett.117.072001}{Phys. Rev. Lett. {\bf
  117} (2016) no.~7, 072001},
\href{http://arxiv.org/abs/1606.00840}{{\tt arXiv:1606.00840 [hep-ph]}}.
%%CITATION = ARXIV:1606.00840;%%.

\bibitem{Campanario:2018ppz}
F.~Campanario, T.~M. Figy, S.~Plätzer, M.~Rauch, P.~Schichtel, and
  M.~Sjödahl, {\em {Stress testing the vector-boson-fusion approximation in
  multijet final states}}.
  \href{http://dx.doi.org/10.1103/PhysRevD.98.033003}{Phys. Rev. D {\bf 98}
  (2018) no.~3, 033003}, \href{http://arxiv.org/abs/1802.09955}{{\tt
  arXiv:1802.09955 [hep-ph]}}.

\bibitem{Han:1992hr}
T.~Han, G.~Valencia, and S.~Willenbrock, {\em {Structure function approach to
  vector boson scattering in p p collisions}}.
  \href{http://dx.doi.org/10.1103/PhysRevLett.69.3274}{Phys. Rev. Lett. {\bf
  69} (1992)  3274--3277},
\href{http://arxiv.org/abs/hep-ph/9206246}{{\tt arXiv:hep-ph/9206246
  [hep-ph]}}.
%%CITATION = HEP-PH/9206246;%%.

\bibitem{Liu:2019tuy}
T.~Liu, K.~Melnikov, and A.~A. Penin, {\em {Nonfactorizable QCD Effects in
  Higgs Boson Production via Vector Boson Fusion}}.
\href{http://arxiv.org/abs/1906.10899}{{\tt arXiv:1906.10899 [hep-ph]}}.
%%CITATION = ARXIV:1906.10899;%%.

\bibitem{Ciccolini:2007ec}
M.~Ciccolini, A.~Denner, and S.~Dittmaier, {\em {Electroweak and QCD
  corrections to Higgs production via vector-boson fusion at the LHC}}.
  \href{http://dx.doi.org/10.1103/PhysRevD.77.013002}{Phys. Rev. {\bf D77}
  (2008)  013002},
\href{http://arxiv.org/abs/0710.4749}{{\tt arXiv:0710.4749 [hep-ph]}}.
%%CITATION = ARXIV:0710.4749;%%.

\bibitem{Harlander:2008xn}
R.~V. Harlander, J.~Vollinga, and M.~M. Weber, {\em {Gluon-Induced Weak Boson
  Fusion}}. \href{http://dx.doi.org/10.1103/PhysRevD.77.053010}{Phys. Rev. {\bf
  D77} (2008)  053010},
\href{http://arxiv.org/abs/0801.3355}{{\tt arXiv:0801.3355 [hep-ph]}}.
%%CITATION = ARXIV:0801.3355;%%.

\bibitem{Andersen:2007mp}
J.~R. Andersen, T.~Binoth, G.~Heinrich, and J.~M. Smillie, {\em {Loop induced
  interference effects in Higgs Boson plus two jet production at the LHC}}.
  \href{http://dx.doi.org/10.1088/1126-6708/2008/02/057}{JHEP {\bf 02} (2008)
  057},
\href{http://arxiv.org/abs/0709.3513}{{\tt arXiv:0709.3513 [hep-ph]}}.
%%CITATION = ARXIV:0709.3513;%%.

\bibitem{Vogt:2004mw}
A.~Vogt, S.~Moch, and J.~A.~M. Vermaseren, {\em {The Three-loop splitting
  functions in QCD: The Singlet case}}.
  \href{http://dx.doi.org/10.1016/j.nuclphysb.2004.04.024}{Nucl. Phys. {\bf
  B691} (2004)  129--181},
\href{http://arxiv.org/abs/hep-ph/0404111}{{\tt arXiv:hep-ph/0404111
  [hep-ph]}}.
%%CITATION = HEP-PH/0404111;%%.

\bibitem{Moch:2004xu}
S.~Moch, J.~A.~M. Vermaseren, and A.~Vogt, {\em {The Longitudinal structure
  function at the third order}}.
  \href{http://dx.doi.org/10.1016/j.physletb.2004.11.063}{Phys. Lett. {\bf
  B606} (2005)  123--129},
\href{http://arxiv.org/abs/hep-ph/0411112}{{\tt arXiv:hep-ph/0411112
  [hep-ph]}}.
%%CITATION = HEP-PH/0411112;%%.

\bibitem{Vermaseren:2005qc}
J.~A.~M. Vermaseren, A.~Vogt, and S.~Moch, {\em {The Third-order QCD
  corrections to deep-inelastic scattering by photon exchange}}.
  \href{http://dx.doi.org/10.1016/j.nuclphysb.2005.06.020}{Nucl. Phys. {\bf
  B724} (2005)  3--182},
\href{http://arxiv.org/abs/hep-ph/0504242}{{\tt arXiv:hep-ph/0504242
  [hep-ph]}}.
%%CITATION = HEP-PH/0504242;%%.

\bibitem{Moch:2007rq}
S.~Moch, M.~Rogal, and A.~Vogt, {\em {Differences between charged-current
  coefficient functions}}.
  \href{http://dx.doi.org/10.1016/j.nuclphysb.2007.09.022}{Nucl. Phys. {\bf
  B790} (2008)  317--335},
\href{http://arxiv.org/abs/0708.3731}{{\tt arXiv:0708.3731 [hep-ph]}}.
%%CITATION = ARXIV:0708.3731;%%.

\bibitem{Buehler:2011ev}
S.~Buehler and C.~Duhr, {\em {CHAPLIN - Complex Harmonic Polylogarithms in
  Fortran}}. \href{http://dx.doi.org/10.1016/j.cpc.2014.05.022}{Comput. Phys.
  Commun. {\bf 185} (2014)  2703--2713},
\href{http://arxiv.org/abs/1106.5739}{{\tt arXiv:1106.5739 [hep-ph]}}.
%%CITATION = ARXIV:1106.5739;%%.

\bibitem{Dobrovolskaya:1990kx}
A.~Dobrovolskaya and V.~Novikov, {\em {On heavy Higgs boson production}}.
\href{http://dx.doi.org/10.1007/BF01559437}{Z. Phys. {\bf C52} (1991)
  427--436}.
%%CITATION = ZEPYA,C52,427;%%.

\bibitem{Cheng:1969ab}
H.~Cheng and T.~T. Wu, {\em {High-energy collision processes in quantum
  electrodynamics. iii}}.
  \href{http://dx.doi.org/10.1103/PhysRev.182.1873}{Phys. Rev. {\bf 182} (1969)
   1873--1898}.

\bibitem{Chang:1969by}
S.-J. Chang and S.-K. Ma, {\em {Multiphoton exchange amplitudes at infinite
  energy}}.
\href{http://dx.doi.org/10.1103/PhysRev.188.2385}{Phys. Rev. {\bf 188} (1969)
  2385--2404}.
%%CITATION = PHRVA,188,2385;%%.

\bibitem{Cheng:1970jk}
H.~Cheng and T.~T. Wu, {\em {Impact factor and exponentiation in high-energy
  scattering processes}}.
\href{http://dx.doi.org/10.1103/PhysRev.186.1611}{Phys. Rev. {\bf 186} (1969)
  1611--1618}.
%%CITATION = PHRVA,186,1611;%%.

\bibitem{Lipatov:1976zz}
L.~N. Lipatov, {\em {Reggeization of the Vector Meson and the Vacuum
  Singularity in Nonabelian Gauge Theories}}. Sov. J. Nucl. Phys. {\bf 23}
  (1976)  338--345.
[Yad. Fiz.23,642(1976)].
%%CITATION = SJNCA,23,338;%%.

\bibitem{Campanario:2013fsa}
F.~Campanario, T.~M. Figy, S.~Plätzer, and M.~Sjödahl, {\em {Electroweak
  Higgs Boson Plus Three Jet Production at Next-to-Leading-Order QCD}}.
  \href{http://dx.doi.org/10.1103/PhysRevLett.111.211802}{Phys. Rev. Lett. {\bf
  111} (2013) no.~21, 211802},
\href{http://arxiv.org/abs/1308.2932}{{\tt arXiv:1308.2932 [hep-ph]}}.
%%CITATION = ARXIV:1308.2932;%%.

\bibitem{Ball:2014uwa}
{\bf NNPDF} Collaboration, R.~D. Ball {\em et al.}, {\em {Parton distributions
  for the LHC Run II}}. \href{http://dx.doi.org/10.1007/JHEP04(2015)040}{JHEP
  {\bf 04} (2015)  040},
\href{http://arxiv.org/abs/1410.8849}{{\tt arXiv:1410.8849 [hep-ph]}}.
%%CITATION = ARXIV:1410.8849;%%.

\bibitem{Buckley:2014ana}
A.~Buckley, J.~Ferrando, S.~Lloyd, K.~Nordström, B.~Page, M.~Rüfenacht,
  M.~Schönherr, and G.~Watt, {\em {LHAPDF6: parton density access in the LHC
  precision era}}. \href{http://dx.doi.org/10.1140/epjc/s10052-015-3318-8}{Eur.
  Phys. J. {\bf C75} (2015)  132},
\href{http://arxiv.org/abs/1412.7420}{{\tt arXiv:1412.7420 [hep-ph]}}.
%%CITATION = ARXIV:1412.7420;%%.

\bibitem{Aad:2015zhl}
{\bf ATLAS, CMS} Collaboration, G.~Aad {\em et al.}, {\em {Combined Measurement
  of the Higgs Boson Mass in $pp$ Collisions at $\sqrt{s}=7$ and 8 TeV with the
  ATLAS and CMS Experiments}}.
  \href{http://dx.doi.org/10.1103/PhysRevLett.114.191803}{Phys. Rev. Lett. {\bf
  114} (2015)  191803},
\href{http://arxiv.org/abs/1503.07589}{{\tt arXiv:1503.07589 [hep-ex]}}.
%%CITATION = ARXIV:1503.07589;%%.

\bibitem{Cacciari:2008gp}
M.~Cacciari, G.~P. Salam, and G.~Soyez, {\em {The Anti-k(t) jet clustering
  algorithm}}. \href{http://dx.doi.org/10.1088/1126-6708/2008/04/063}{JHEP {\bf
  0804} (2008)  063},
\href{http://arxiv.org/abs/0802.1189}{{\tt arXiv:0802.1189 [hep-ph]}}.
%%CITATION = ARXIV:0802.1189;%%.

\bibitem{Cacciari:2011ma}
M.~Cacciari, G.~P. Salam, and G.~Soyez, {\em {FastJet User Manual}}.
  \href{http://dx.doi.org/10.1140/epjc/s10052-012-1896-2}{Eur.Phys.J. {\bf C72}
  (2012)  1896},
\href{http://arxiv.org/abs/1111.6097}{{\tt arXiv:1111.6097 [hep-ph]}}.
%%CITATION = ARXIV:1111.6097;%%.

\bibitem{Salam:2008qg}
G.~P. Salam and J.~Rojo, {\em {A Higher Order Perturbative Parton Evolution
  Toolkit (HOPPET)}}.
  \href{http://dx.doi.org/10.1016/j.cpc.2008.08.010}{Comput. Phys. Commun. {\bf
  180} (2009)  120--156},
\href{http://arxiv.org/abs/0804.3755}{{\tt arXiv:0804.3755 [hep-ph]}}.
%%CITATION = ARXIV:0804.3755;%%.

\bibitem{Alioli:2010xd}
S.~Alioli, P.~Nason, C.~Oleari, and E.~Re, {\em {A general framework for
  implementing NLO calculations in shower Monte Carlo programs: the POWHEG
  BOX}}. \href{http://dx.doi.org/10.1007/JHEP06(2010)043}{JHEP {\bf 1006}
  (2010)  043},
\href{http://arxiv.org/abs/1002.2581}{{\tt arXiv:1002.2581 [hep-ph]}}.
%%CITATION = ARXIV:1002.2581;%%.

\bibitem{Nason:2009ai}
P.~Nason and C.~Oleari, {\em {NLO Higgs boson production via vector-boson
  fusion matched with shower in POWHEG}}.
  \href{http://dx.doi.org/10.1007/JHEP02(2010)037}{JHEP {\bf 02} (2010)  037},
\href{http://arxiv.org/abs/0911.5299}{{\tt arXiv:0911.5299 [hep-ph]}}.
%%CITATION = ARXIV:0911.5299;%%.

\bibitem{Khachatryan:2015bnx}
{\bf CMS} Collaboration, V.~Khachatryan {\em et al.}, {\em {Search for the
  standard model Higgs boson produced through vector boson fusion and decaying
  to $b \overline{b}$}}.
  \href{http://dx.doi.org/10.1103/PhysRevD.92.032008}{Phys. Rev. {\bf D92}
  (2015) no.~3, 032008},
\href{http://arxiv.org/abs/1506.01010}{{\tt arXiv:1506.01010 [hep-ex]}}.
%%CITATION = ARXIV:1506.01010;%%.

\bibitem{Aaboud:2018gay}
{\bf ATLAS} Collaboration, M.~Aaboud {\em et al.}, {\em {Search for Higgs
  bosons produced via vector-boson fusion and decaying into bottom quark pairs
  in $\sqrt{s} = 13$ $\mathrm{TeV}$ $pp$ collisions with the ATLAS detector}}.
  \href{http://dx.doi.org/10.1103/PhysRevD.98.052003}{Phys. Rev. {\bf D98}
  (2018) no.~5, 052003},
\href{http://arxiv.org/abs/1807.08639}{{\tt arXiv:1807.08639 [hep-ex]}}.
%%CITATION = ARXIV:1807.08639;%%.

\bibitem{Rauch:2017cfu}
M.~Rauch and D.~Zeppenfeld, {\em {Jet clustering dependence of Higgs boson
  production in vector-boson fusion}}.
  \href{http://dx.doi.org/10.1103/PhysRevD.95.114015}{Phys. Rev. {\bf D95}
  (2017) no.~11, 114015},
\href{http://arxiv.org/abs/1703.05676}{{\tt arXiv:1703.05676 [hep-ph]}}.
%%CITATION = ARXIV:1703.05676;%%.

\bibitem{Dreyer:2018rfu}
F.~A. Dreyer and A.~Karlberg, {\em {Fully differential Vector-Boson Fusion
  Higgs Pair Production at Next-to-Next-to-Leading Order}}.
  \href{http://dx.doi.org/10.1103/PhysRevD.99.074028}{Phys. Rev. {\bf D99}
  (2019) no.~7, 074028},
\href{http://arxiv.org/abs/1811.07918}{{\tt arXiv:1811.07918 [hep-ph]}}.
%%CITATION = ARXIV:1811.07918;%%.

\bibitem{Dreyer:2018qbw}
F.~A. Dreyer and A.~Karlberg, {\em {Vector-Boson Fusion Higgs Pair Production
  at N$^3$LO}}. \href{http://dx.doi.org/10.1103/PhysRevD.98.114016}{Phys. Rev.
  {\bf D98} (2018) no.~11, 114016},
\href{http://arxiv.org/abs/1811.07906}{{\tt arXiv:1811.07906 [hep-ph]}}.
%%CITATION = ARXIV:1811.07906;%%.

\bibitem{provbfh}
proVBFH.
\newblock {\url{http://provbfh.hepforge.org/}}.

\end{thebibliography}\endgroup
% \bibliographystyle{unsrtnat}
% \begin{thebibliography}{99}

% \end{thebibliography}

\end{document}